\documentclass{aa}  

\usepackage{graphicx}
\usepackage{txfonts}
\usepackage{hyperref}
\usepackage[]{natbib}
\bibpunct{(}{)}{;}{a}{}{,}  
\usepackage{rotating}

\newcommand{\ratio} {N({\rm H}_2) / I_{\rm CO(1-0)}}
\newcommand{\ratioo} {N({\rm H}_2) / I_{\rm CO}}
\newcommand{\ratiot} {N({\rm H}_2) / I_{\rm CO(2-1)}}
\newcommand{\kms}   {{\rm \  km \  s^{-1}}}
\newcommand{\Xunit} {\,{\rm cm^{-2}/(K\kms)}}

\begin{document} 

\title{The IRAM M~33 CO(2--1) Survey}
		\subtitle{A complete census of molecular gas out to 7~kpc}
		
  \author{
    C.~Druard\inst{1,2}
    \and
    J.~Braine\inst{1,2}
    \and
    K.~F.~Schuster\inst{3}
    \and
    N.~Schneider\inst{1,2}
    \and
    P.~Gratier\inst{1,2}
    \and
    S.~Bontemps\inst{1,2}
    \and
    M.~Boquien\inst{4}
    \and
    F.~Combes\inst{5}
    \and
    E.~Corbelli\inst{6}
    \and
    C.~Henkel\inst{7,8}
    \and
    F.~Herpin\inst{1,2}
     \and
    C.~Kramer\inst{9}
    \and
    F.~van~der~Tak\inst{10,11}
    \and
    P.~van~der~Werf\inst{12}
  }

  \institute{
    Univ. Bordeaux, LAB, UMR 5804, F-33270, Floirac, France \\ \email{druard@obs.u-bordeaux1.fr}
    \and
    CNRS, LAB, UMR 5804, F-33270, Floirac, France
    \and
    IRAM, 300 rue de la Piscine, F-38406 St. Martin d'H\`eres, France
    \and
    Aix-Marseille Université, CNRS, LAM (Laboratoire d’Astrophysique de Marseille) UMR 7326, 13388 Marseille, France
    \and
    Observatoire de Paris, LERMA, CNRS: UMR8112, 61 Av. de l’Observatoire, 75014 Paris, France
    \and
    Osservatorio Astrofisico di Arcetri – INAF, Largo E. Fermi 5, 50125 Firenze, Italy
    \and
    Max-Planck-Institut für Radioastronomie, Auf dem Hügel 69, 53121 Bonn, Germany
    \and
    Astron. Dept., King Abdulaziz University, P.O. Box 80203, Jeddah 21589, Saudi Arabia
    \and
    Instituto Radioastronomía Milimétrica (IRAM), Av. Divina Pastora 7, Nucleo Central, 18012 Granada, Spain
    \and
    SRON Netherlands Institute for Space Research, Landleven 12, 9747 AD Groningen, The Netherlands
    \and
    Kapteyn Astronomical Institute, University of Groningen, The Netherlands
    \and
    Leiden Observatory, Leiden University, P.O. Box 9513, NL-2300 RA Leiden, The Netherlands
  }
             
\date{}

	\abstract{
    In order to study the interstellar medium and the interplay between the atomic and molecular components in a low-metallicity environment, we present a complete high angular and spectral resolution map and position-position-velocity data cube of the \element[ ][12]{CO}(J=2--1) emission from the Local Group galaxy Messier 33. Its metallicity is roughly half-solar, such that we can compare its interstellar medium with that of the Milky Way with the main changes being the metallicity and the gas mass fraction. The data have a 12$\arcsec$ angular resolution ($\sim$50~pc) with a spectral resolution of 2.6$\kms$ and a mean and median noise level of 20~mK per channel in antenna temperature. A radial cut along the major axis was also observed in the \element[ ][12]{CO}(J=1--0) line.
    The CO data cube and integrated intensity map are optimal when using \ion{H}{i} data to define the baseline window and the velocities over which the CO emission is integrated. 
    Great care was taken when building these maps, testing different windowing and baseline options and investigating the effect of error beam pickup. The total CO(2--1) luminosity is 2.8~$\times$~10$^{7}$K$\kms$pc$^{2}$, following the spiral arms in the inner disk, with an average decrease in intensity approximately following an exponential disk with a scale length of 2.1~kpc. There is no clear variation in the CO($\frac{2-1}{1-0}$) intensity ratio with radius and the average value is roughly 0.8. The total molecular gas mass is estimated, using a $\ratio$ = 4~$\times$~10$^{20}$ $\Xunit$ conversion factor, to be 3.1~$\times$~10$^8$~M$_\sun$, including Helium. 
    The CO spectra in the cube were shifted to zero velocity by subtracting the velocity of the HI peak from the CO spectra. Stacking these spectra over the whole disk yields a CO line with a half-power width of 12.4$\kms$. Hence, the velocity dispersion between the atomic and molecular components is extremely low, independently justifying the use of the HI line in building our maps. Stacking the spectra in concentric rings shows that the CO linewidth and possibly the CO-HI velocity dispersion decrease in the outer disk. The error beam pickup could produce the weak CO emission apparently from regions in which the \ion{H}{i} line peak does not reach 10~K, such that no CO is actually detected in these regions. 
    Using the CO(2--1) emission to trace the molecular gas, the probability distribution function of the H$_2$ column density shows an excess at high column density above a log-normal distribution.
  }
  

	\maketitle   

\section{Introduction}

  The process of phase transition from the important mass reservoir of atomic hydrogen in galactic discs and the dense star-forming molecular phase is a matter of intense research. The Local Group spiral galaxy M~33, is ideally suited for a study of its molecular gas content and the dependence of the star formation characteristics on the physical conditions across the spatially resolved galactic disk. At a distance of 840~kpc, the Triangulum galaxy is near enough to resolve individual Giant Molecular Clouds (GMCs) with large radio telescopes such as the IRAM 30 meter antenna. M~33 is a small but classical spiral disk with a mass roughly 10\% \citep{Corbelli2003} that of the Milky Way and, unlike Andromeda and our galaxy (the other two local group spirals), it is inclined such that positions and velocities can be determined with no ambiguity. M~33 is chemically young, with a high gas mass fraction, and as such represents a different environment in which to study cloud and star formation with respect to the Milky Way. As the average metallicity is subsolar by roughly a factor 2, M~33 represents a stepping stone towards much lower metallicity objects and is the nearest late-type galaxy with a well observed metallicity gradient \citep{Magrini2007,Magrini2010,Rosolowsky2008}. 

  In this work we present a high-sensitivity and high-resolution survey of the CO(2--1) emission from M~33, covering the full optical disk. M~33 has an inclination of $i=$ 56$\degr$ that makes the position of the clouds in the disk well defined, in contrast to e.g. M~31. 
  The observations of the CO(2--1) line presented here have an angular resolution of 12$\arcsec$, corresponding to 49~pc at a distance of 840~kpc for M~33. Single-dish maps do not suffer from missing flux problems often present in interferometric data, an important asset for the understanding of the entire molecular phase in galactic disks. Relevant comparison studies are \citet{Engargiola2003}, \cite{Rosolowsky2007} and \citet{Tosaki2011} for M~33 at 13$\arcsec$, 15$\arcsec$ and 19.3$\arcsec$ resolution, PAWS \citet{Schinnerer2013} for M~51 and \citet{Kawamura2009} and \citet{Wong2011} for the Large Magellanic Cloud. None of these works reach a brightness sensitivity or a coverage comparable to the observations presented here.
  
\begin{figure*}[p]
    \centering
    \includegraphics[width=16cm]{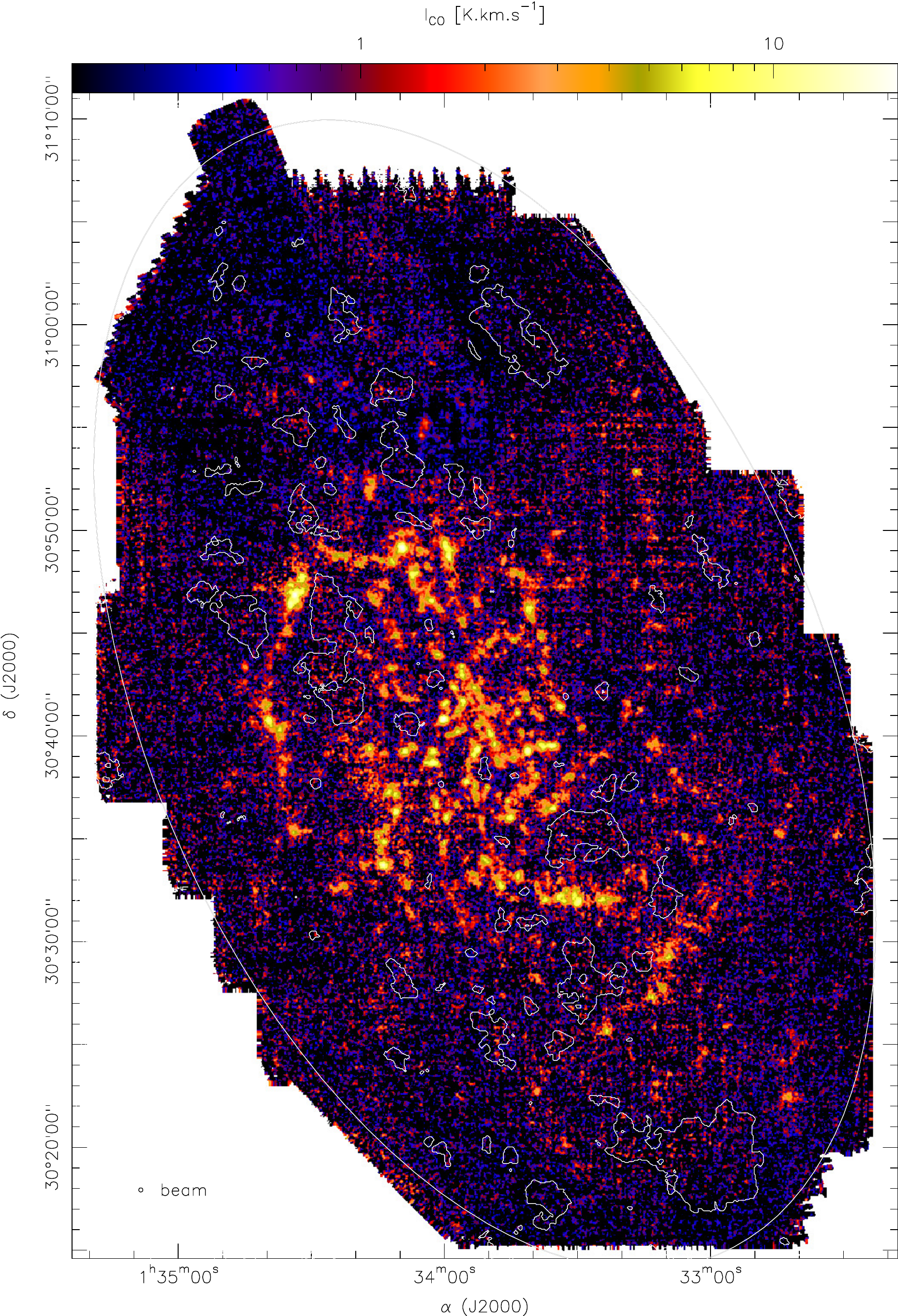}
    \caption{CO(2--1) integrated intensity map in K$\kms$, expressed in the main beam temperature scale and computed as described in Sect.~\ref{sec:iconoHI}. The contours show \ion{H}{i} -poor regions where the \ion{H}{i}  line does not reach 10~K. The beam size is shown in the lower left corner of the figure. The white ellipse represents a 7.2~kpc radius from the center.}
    \label{fig:ico0K}
  \end{figure*}

  The data reduction, leading to a data cube, and the different methods of obtaining integrated intensity maps are discussed in detail.  These are the main data products of the survey (see e.g. the integrated intensity map in Fig.~\ref{fig:ico0K}) and will be made publicly available on the IRAM Large Program archival web page\footnote{\url{http://www.iram-institute.org/EN/content-page-240-7-158-240-0-0.html}}. This naturally leads to an investigation of the error-beam pickup, something which has not been done in other studies of M~33.  The error beam pickup is relevant to assess the reality of low-level diffuse CO emission; in particular, as we use the  \ion{H}{i} line information in the reduction process, we would like to determine whether \ion{H}{i}-free regions are really CO-free. Finally, we compare the CO and \ion{H}{i} line profiles and velocities over the disk, showing that M~33 is a dynamically very cool disk.

  Extensive ancillary data are available for M~33, particularly the \ion{H}{i} \citep{Gratier2010a}, multiband Spitzer data \citep{Tabatabaei2007,Verley2007}, and Herschel observations from the HerM33es project \citep{Kramer2010,Boquien2011}.
  
  Companion papers are in preparation on ($i$) the $\ratio$ factor, using the Herschel and \ion{H}{i} data and including a possible hidden H$_2$ component from PDR (Photon-Dominated Regions) layers which are not traced by CO (Gratier et al., in prep.); ($ii$) a detailed study of the molecular cloud population (Druard et al., in prep.); ($iii$) a comparison with models \citep[e.g.][]{Blitz2006,Krumholz2008, Gnedin2009} which attempt to explain where molecular clouds form (Braine et al., in prep.); ($iv$) a study of the probability distribution functions of various tracers (dust, CO, \ion{H}{I}) (Schneider et al., in prep.). ($v$) and a comparison of GMC and Young Stellar Cluster positions to deduce a time scale for the GMC life cycle (Corbelli et al., in prep.).\\

\section{Data and Observations}

  \subsection{The complete CO(2--1) dataset}
\begin{figure}
    \centering
    \includegraphics[width=9cm]{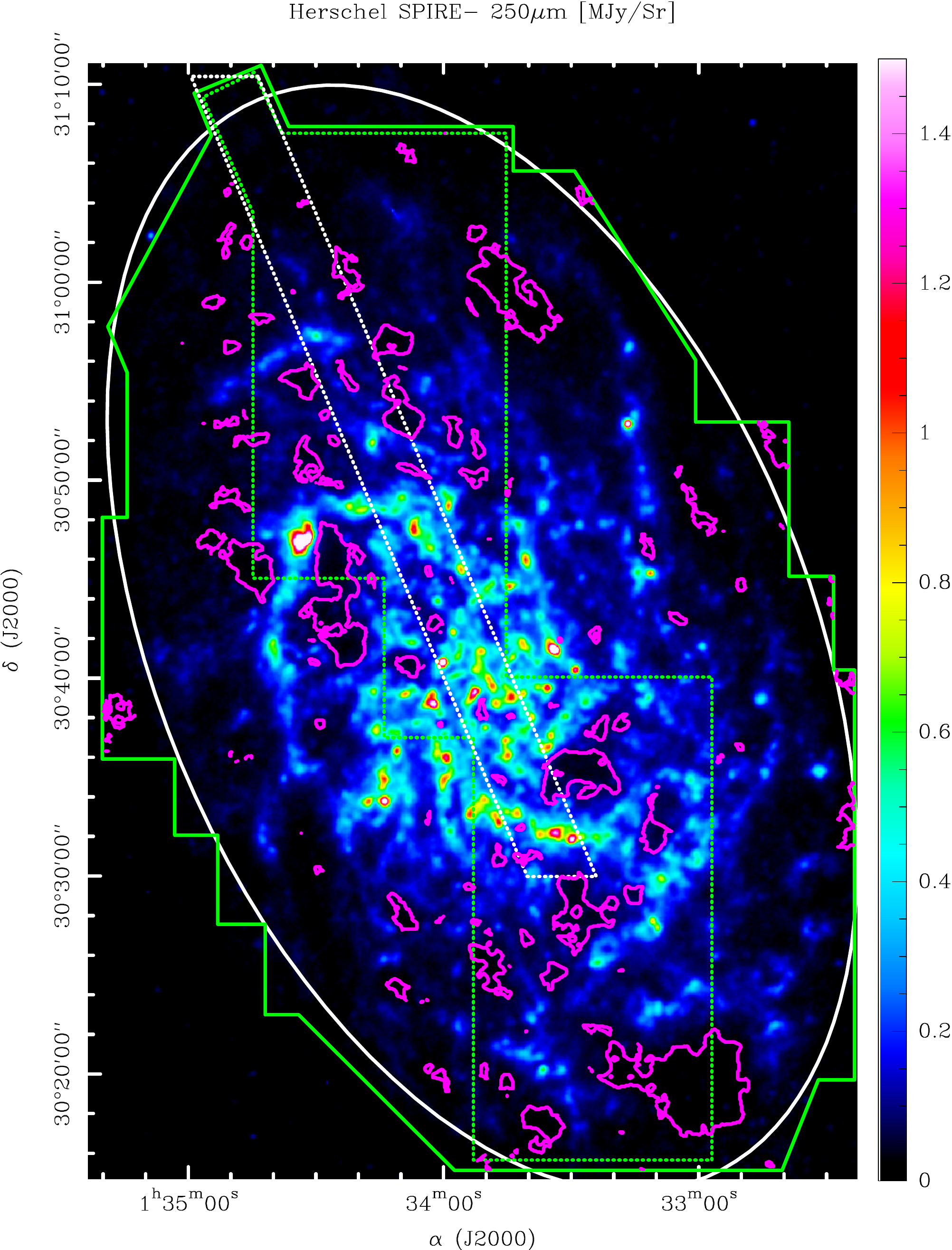}
    \caption{The Triangulum galaxy M~33. This image shows in plain green the edges of the coverage of the CO(2--1) survey on top of a Herschel 250 $\mu$m map \citep{Xilouris2012} that traces cold dust. The green dotted limits correspond to the previous processed area in \citet{Gratier2010a}. The 2$\arcmin$~$\times$~40$\arcmin$ wide HerM33es strip, in white dots, is also the region here the CO(1--0) transition has been observed with the 30~m telescope. The purple limited areas (white lines in Fig.~\ref{fig:ico0K}) are the \ion{H}{i}-poor regions where we detect no \ion{H}{i} above 10~K and inside which we can see that there is no strong 250 $\mu$m emission. The white ellipse represents a 7.2~kpc radius from the center.}
    \label{fig:coverage}
  \end{figure}

  M~33 was observed in the CO(2--1) line at 230.538~GHz with the HEterodyne Receiver Array \citep[HERA,][]{HERA} on the 30 meter telescope of the Institut de RadioAstronomie Millim\'etrique (IRAM) on Pico Veleta in southern Spain\footnote{IRAM is supported by INSU/CNRS (France), MPG (Germany) and IGN (Spain).}. These observations are a follow-up to \citet{Gardan2007} and \citet{Gratier2010a} which covered only part of M~33 (see Fig.~\ref{fig:coverage}). Observing parameters (see Table \ref{tab:obsparam}) and procedures have been maintained throughout the observing period (2005--2012). 

  \begin{table}
    \caption{\element[][12]{CO}(2--1) observational parameters}
    \label{tab:obsparam}
    \centering
    \begin{tabular}{l c}
      \hline
      \multicolumn{2}{c}{M~33 characteristics} \\
      \hline\hline
      $\alpha$ (J2000) & 1\textsuperscript{h}33\textsuperscript{min}50.9\textsuperscript{s} \\
      $\delta$ (J2000) & +30\degr39\arcmin35.80\arcsec \\
      type & SA(s)cd (1) \\
      LSR velocity & -170$\kms$ \\
      position angle & 22,5\degr (2)\\
      inclination & 56\degr (3)\\
      distance & 840~kpc (4)\\
      \hline
      \multicolumn{2}{c}{telescope parameters} \\
      \hline\hline
      beam size & 10.7\arcsec \\
      number of independent dumps & 20 648 846 \\
      channel width & 2.6$\kms$ - 2~MHz \\
      size of the map & $2400\arcsec~\times~3400\arcsec$ \\
      mean RMS noise & 20.33~mK \\
      \hline
    \end{tabular}
    \tablebib{(1)~\citet{Vaucouleurs}; (2) \citet{Paturel2003}; (3) \citet{Regan1994}; (4) \citet{Galleti2004};}
  \end{table}
  
  The multi-beam 230~GHz receiver HERA consists of a 3~$\times$~3 array of dual polarization pixels \citep{HERA}. HERA was used in the On-The-Fly scanning mode in which spectra are taken rapidly while the telescope is moving continuously with a constant angular velocity. Integration times for individual spectra, or dumps, are typically 0.5 seconds with a scanning speed of 6$\arcsec$ per second. Reference spectra towards positions where no emission is expected are taken regularly. These "OFF" spectra are considerably longer than the individual dumps so the subtraction of the reference adds essentially no noise.  We used the WILMA autocorrelator which measures spectra with a 2.6$\kms$ (2~MHz) channel spacing. A second autocorrelator (VESPA) with 1.25~MHz channel spacing was also used starting in 2007 \citep{Gardan2007}, but these data have not been used as the original northern region was not observed at that spectral resolution.

  M~33 was divided up into fields which could be covered in a convenient time, typically about 30 to 60 minutes, about half the time between pointings. These areas were mostly observed in strips parallel to right-ascension or declination while some edges of the map were done with slanted strips (see Fig.~\ref{fig:coverage}). Between two coverages of the same region, the array was rotated by 90$\degr$, 180$\degr$ or 270$\degr$ by means of the K-mirror derotator. This keeps the geometry of the 3~$\times$~3 pixel array unchanged but avoids having the same pixels cover exactly the same positions as the differences in receiver temperatures from one pixel to another lead to strips of increased noise levels.  
  
  All the data presented here are in main beam temperature scale, converted with a forward efficiency of $F_{\rm eff} = 0.92$ and a main beam efficiency of $B_{\rm eff} = 0.56$. These are {\it not} the same as used by \citet{Gardan2007} and \citet{Gratier2010a} because updated telescope efficiency measurements covering our observing period have recently become available\footnote{\url{http://www.iram-institute.org/medias/uploads/eb2013-v8.2.pdf}}. The HERA efficiencies published by \citet{HERA} were deduced from measurements made in 2001, prior to the major surface improvements in 2002. The efficiencies and error beam are essentially a function of the telescope but the average coupling of the multibeam receivers is somewhat poorer.  Therefore, we use efficiency measurements of the HERA multibeam with measurements of the efficiencies of the single-beam receivers to estimate the difference in coupling and the post-2002 measurements of the single-beam (EMIR) beam efficiency. The forward efficiency is assumed to be the same for the single and multi-beam receivers. All values are for a 230~GHz observing frequency.

  Thus, in 2001, the HERA $B_{\rm eff}$ was 0.49 and, for the A230 and B230 receivers respectively 0.52 and 0.50. We then estimate the HERA $B_{\rm eff} = 0.49/0.51 * 0.59 \approx 0.56$ where 0.51 is the average for the A230 and B230 receivers and 0.59 is the current value for EMIR at 230~GHz. Our adopted $B_{\rm eff}$ and $F_{\rm eff}$ values for the post-2002 period are then different for all the observations presented here than from earlier works \citep{Gardan2007,Gratier2010a}: it is important to take into account this change.

  More than 2~$\times$~10$^7$ spectra were acquired in over 400 hours covering a field of view of 55$\arcmin$~$\times$~40$\arcmin$ amounting to about 400~Gb of data when including both backends.

  At the beginning of each observing session, we pointed towards a strong CO(2--1) source located in M~33 at $\alpha =$ 1\textsuperscript{h}34\textsuperscript{m}09.4\textsuperscript{s}, $\delta =$ +30\degr 49\arcmin 06\arcsec, (J2000) in order to check if the system is correctly tuned and to check the reliability of the incoming spectra before acquiring more data. A spectrum of this position has already been given in Fig.~17 of \citet{Gratier2010a}.

  \subsection{Reduction process}
  \label{sec:reduction}
  The data reduction was carried out with the GILDAS packages CLASS and GREG\footnote{\url{http://www.iram.fr/IRAMFR/GILDAS}}. The pipeline used to reduce the CO(2--1) data was adapted from the one already used in the reduction of the first subsets of data \citep[see]{Gardan2007,Gratier2010a}.

  The WILMA backend was attached to the receiver as to center the spectrum in one half of the spectrometer, which is continuous, therefore excluding platforming (a difference in continuum level) within the velocity range of M~33 (roughly -270 to -90$\kms$). We kept velocities from -400$\kms$ to 0$\kms$ (LSR reference frame) with a channel width of 2.6$\kms$. Because spectra are taken with a 0.5 sec integration time, the CO(2--1) line is invisible in individual spectra. The first step is to fit a constant continuum level (zeroth order polynomial baseline) to each spectrum with no baseline window. Thus the RMS noise of each spectrum is computed around this fit and then compared to the "theoretical noise" given by the following relation :
  \begin{equation}
  \centering
      \sigma_{\mathrm{theo}} \approx \frac{T_{\mathrm{sys}}}{\sqrt{\Delta_{\nu}t}}
  \end{equation}
  with the system temperature $T_{sys}$ in Kelvins, the channel width $\Delta_{\nu}$ in Hz, $t$ the integration time in seconds. The spectra presenting a noise level higher than 1.1\,$\sigma_{\mathrm{theo}}$ are filtered out; this corresponds to nearly 11\% of our dataset.
   
  The disk of M~33 was divided into fields that can be observed in the time between two pointings. Each field was observed multiple times to reduce the RMS noise level and thus enhance the signal to noise ratio. So, as a consequence of the many passes, each position in the sky is associated to multiple spectra observed at similar but typically not identical positions. The typical spacing between spectra is about 3$\arcsec$ so the dataset is oversampled. A large table of the individual spectra is made and the spectra are combined to obtain a regularly gridded position-position-velocity data cube, setting the resolution to 12$\arcsec$ and the pixel size to 3$\arcsec$. We use the XY\_MAP procedure of GILDAS to convolve the spectra into a data cube. The convolution kernel used in the gridding process is a Gaussian 3 times the size of the beam FWHM. We can then convert the cube to a sample of regularly gridded spectra with a much lower noise level than the individual initial 0.5 second integration time spectra.
  
  Up to this stage, apart from a constant continuum level, no baseline has been subtracted. Although most data were taken under good conditions and severe baseline problems were eliminated through a comparison with $\sigma_{\mathrm{theo}}$, the data can be improved by fitting a baseline, typically a polynomial. We compared the noise levels obtained from subtracting polynomial baselines of order 1 to 5. When fitting a baseline, a line window is defined and a polynomial is fit to the remaining channels. A low order polynomial guarantees that no major oscillation will occur within the line window. However, some spectra may require a higher order polynomial to fit the baseline fluctuations. The goal is to subtract the lowest order polynomial which fits the baselines well.

  Fitting polynomial baselines of order 1 to 5, we find that the RMS noise decreases with increasing order but only very slowly, about 0.2~mK/channel ($\sim$1\% of the overall noise) for each increase in baseline order. Our preferred baseline is 3\textsuperscript{rd} order for two reasons: ($i$) the improvement between orders 2 and 3 is greater than between the other increments in baseline order and ($ii$) several regions of the cube remain noisy when a first or second order polynomial is subtracted but the problem is solved by using a third order polynomial and little further improvement appears when going to a fourth or fifth order baseline. Thus, we subtracted a 3\textsuperscript{rd} order baseline from all spectra as the lowest order polynomial allowing the baselines to be consistently well fit.
  Other tests compared the flux in the moment zero (integrated intensity) maps associated with the cubes reduced with different baseline order fits (see Sect.~\ref{sec:icocomput}).
  For a baseline order below 3, the fit appears to miss some of the CO emission but increasing the order does not increase the total emission.
    
  To fit the baseline, the line is excluded from the fit by using velocity windows based on the \ion{H}{i} emission maps \citep{Gratier2010a}. These windows are computed making the assumption that molecular gas mostly forms from atomic gas in M~33 so that H$_2$ will not be present in M~33 at velocities where there is no \ion{H}{i}. Extensive tests were done by using windows based on the M~33 synthetic rotation curve (built from a tilted ring model combined with a deconvolution of the \ion{H}{I} Arecibo data and given by formula 4 in \citet{Corbelli1997}) or on the \ion{H}{i} peak velocity, each of them with variations of the width of the window. We concluded that the \ion{H}{i}-based windows are the ones that yield the lowest mean RMS noise level. A further indication that this choice is appropriate can be found in Sect.~\ref{sec:recenter}. 

\begin{figure}
    \centering
    \includegraphics[width=9cm]{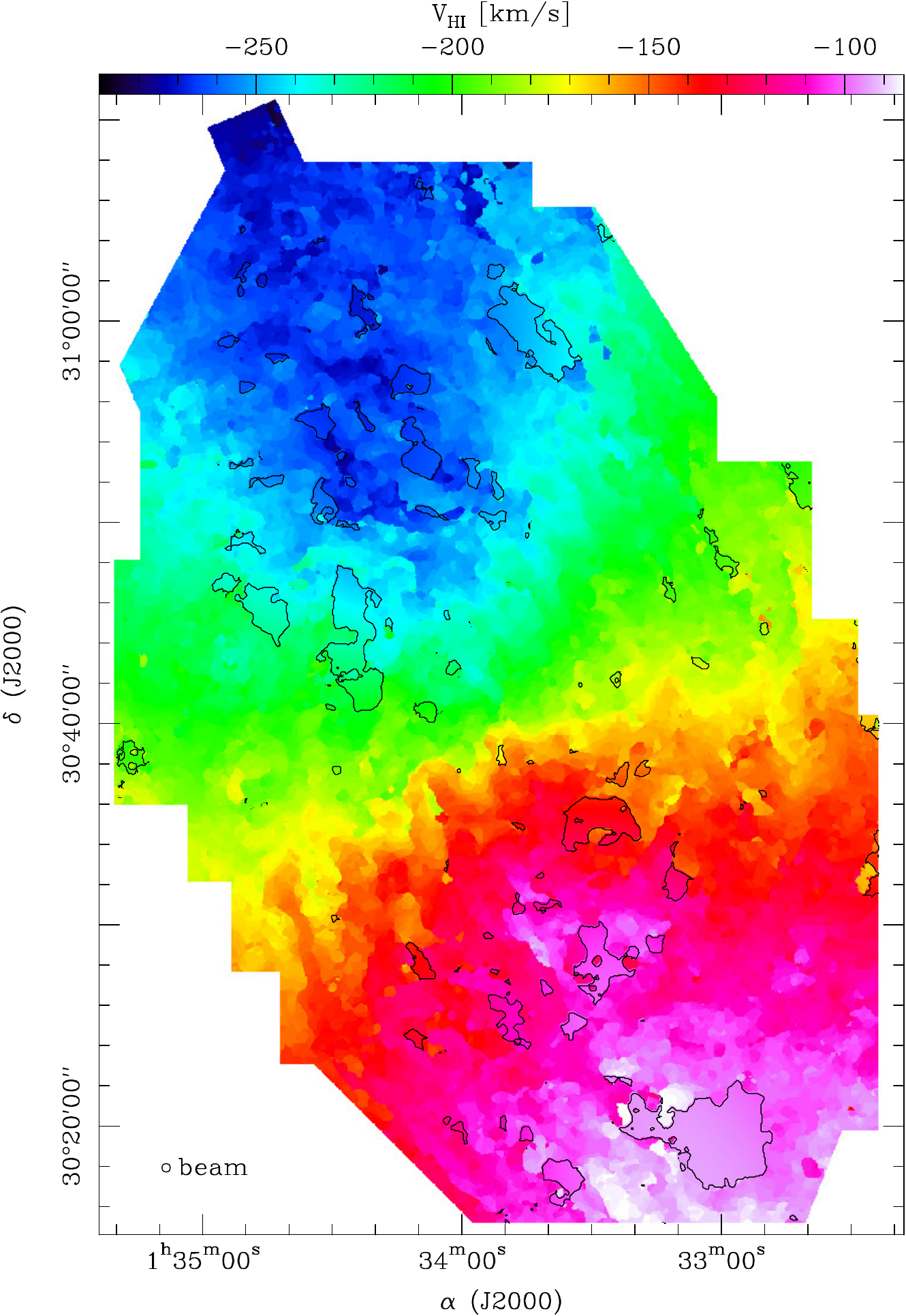}\\
    \caption{\ion{H}{i} peak velocity map in $\kms$. Areas limited in black are regions where there is no 4\,$\sigma$ detection of \ion{H}{i}. In these regions the map is completed with the analytical rotation curve (see Sect.~\ref{sec:reduction}). The 25$\arcsec$~$\times$~25$\arcsec$ beam size for \ion{H}{i} data is shown in the lower left corner.}
    \label{fig:veloHI}
  \end{figure}

\begin{figure}
    \centering
    \includegraphics[width=9cm]{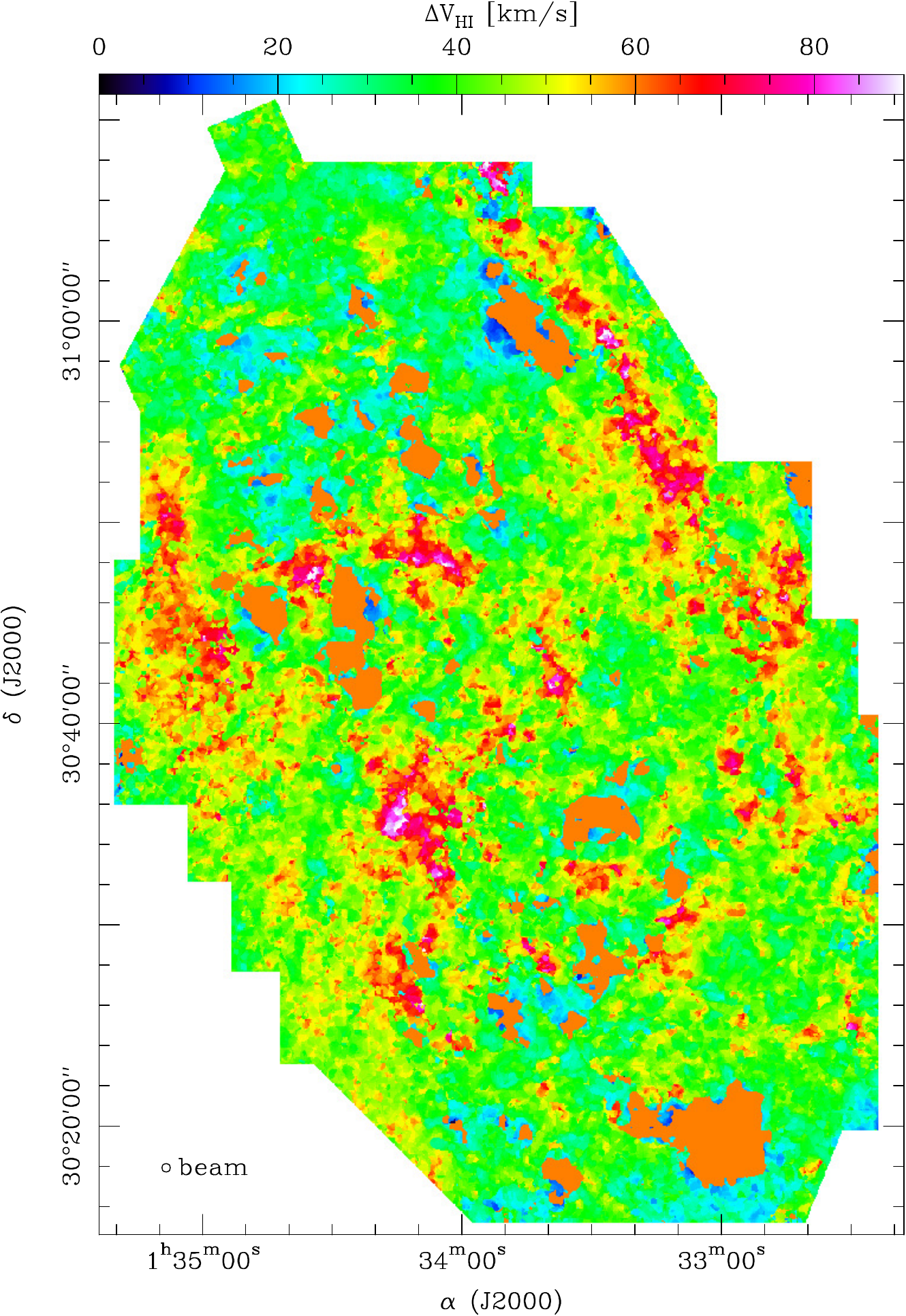}\\
    \caption{Map of the window width used in the reduction process, in $\kms$, corresponding to the \ion{H}{i} based mask used in the baselining process and in the integrated intensity maps computing. Orange regions show areas where no \ion{H}{i} is detected, and where the window is thus equal to 60$\kms$. The 25$\arcsec$~$\times$~25$\arcsec$ beam size for \ion{H}{i} data is shown on the lower left corner.}
    \label{fig:deltaV}
  \end{figure}

  The masking method we used required two types of data: the analytic rotation curve \citep[][Eq. 4]{Corbelli1997} and the \ion{H}{i} data of M~33 already presented and described in \citet{Gratier2010a}. We used the VLA \ion{H}{i} at 25$\arcsec$~$\times$~25$\arcsec$ resolution and channel sampling of 1.27$\kms$ because the noise level is very low -- only 2.6~K on average -- while keeping an angular resolution comparable to that of GMCs or \ion{H}{i} clouds. The \ion{H}{i} data is also available via the Centre de Données de Strasbourg (CDS). For each position in the CO data cube, the velocity limits are calculated by locating the peak \ion{H}{i} line temperature and then going down to the first $\le$ 0~K channel on each side of this peak. The lower and upper limits of the CO window are the velocities of these \ion{H}{i}-free channels. In order to avoid false detections, only the region within $\pm 30\kms$ of the analytical rotation curve is searched for \ion{H}{i}. When no emission above 10K (roughly a 4\,$\sigma$ detection) is present, the line window is taken to be 60$\kms$ centered on the \citet{Corbelli1997} rotation curve. The line windows defined in this way are used to subtract baselines from the CO spectra. The peak \ion{H}{i} velocities are shown in Fig.~\ref{fig:veloHI}. This map is not as smooth as an analytical velocity field, but it follows the \ion{H}{i} velocity variations much more closely. A large velocity shift ($\ge 10\kms$) between two neighboring pixels usually coincides with double peaked spectra where nearby pixels are dominated by different velocity components. For example, the \ion{H}{I} velocity detected around $\alpha =$ 1\textsuperscript{h}34\textsuperscript{min}03\textsuperscript{s}, $\delta =$+30\degr39\arcmin35.80\arcsec (J2000) can be -134$\kms$ for one pixel and -160$\kms$ for its neighbor. Figure \ref{fig:deltaV} shows the width of the line window over the disk of M~33.  

\begin{figure*}
    \centering
    \includegraphics[width= 9cm]{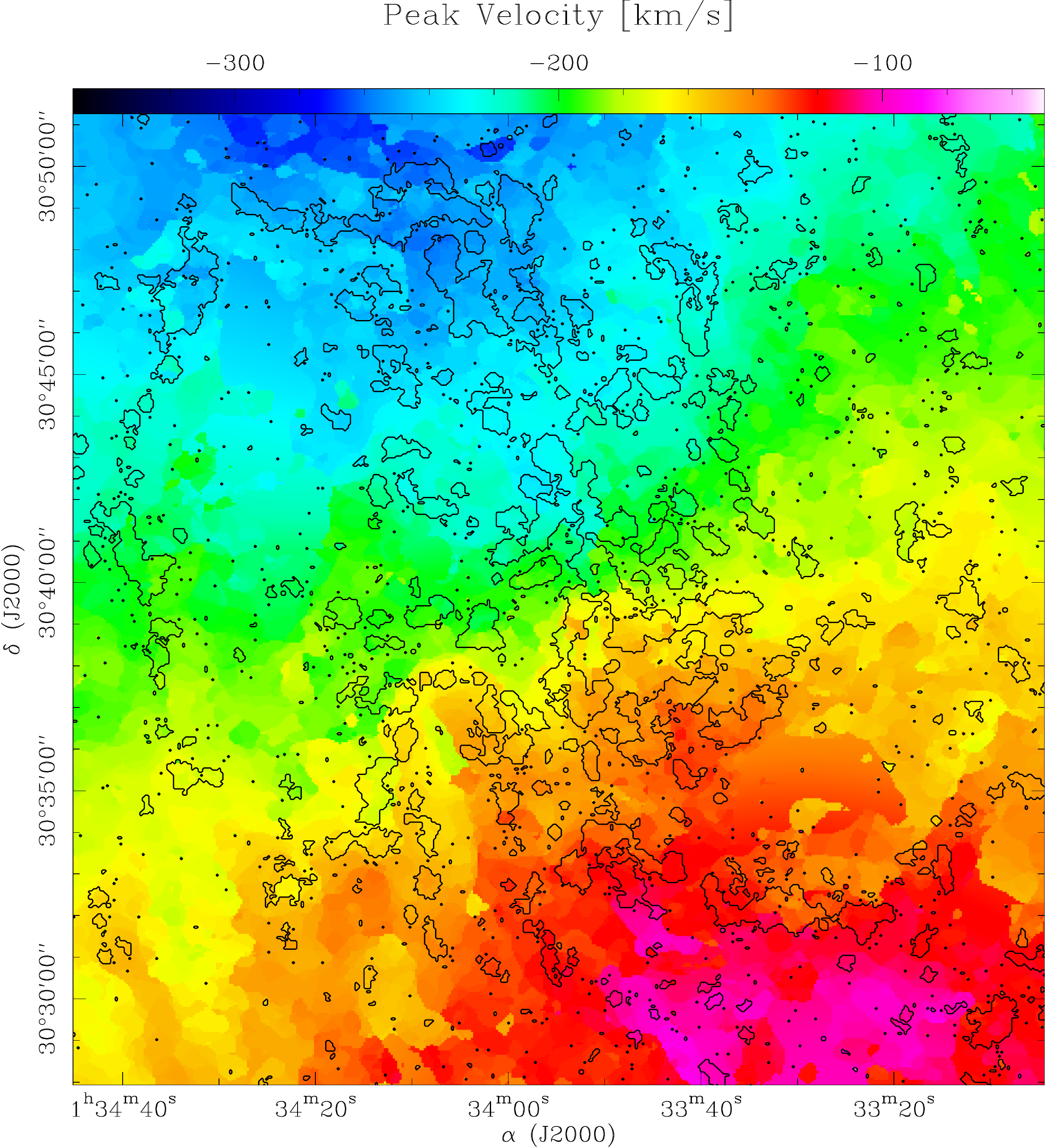}
    \includegraphics[width= 9cm]{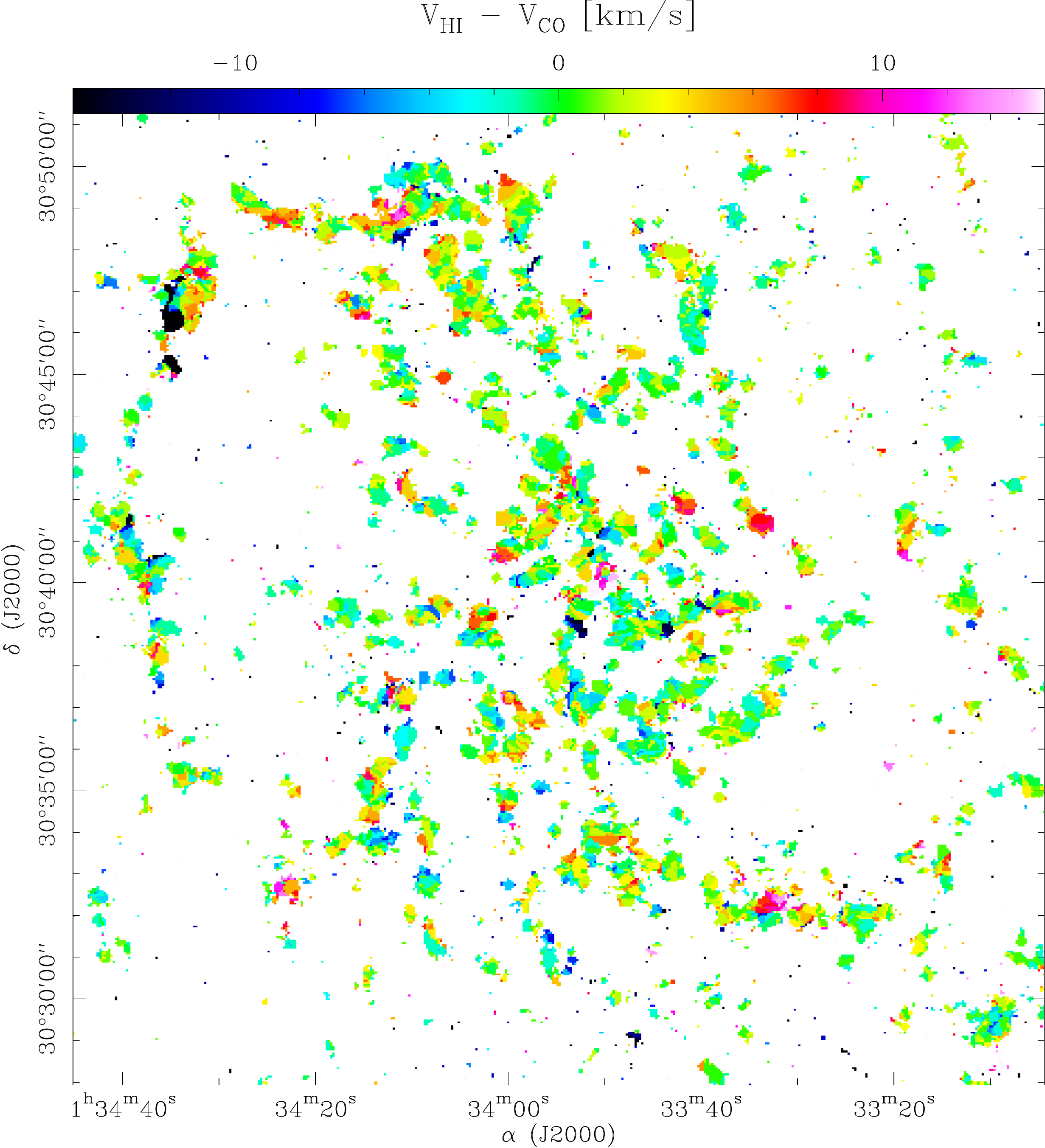}
    \caption{\emph{Left panel.} CO velocity field (black contoured) on top of \ion{H}{I} velocity field map. Inside the black contours the colors represent the CO velocity, while outside they correspond to the \ion{H}{I} velocity. The CO velocity is determined by the velocity of the line peak for 4\,$\sigma$ detections. This map shows that the CO and the \ion{H}{I} peak line temperature are detected at very similar velocities. \emph{Right panel.} \ion{H}{I}-CO velocity shift for the regions where CO velocity is determined. The shift is maximal in NGC604 ($\alpha =$ 1\textsuperscript{h}34\textsuperscript{min}32\textsuperscript{s}, $\delta =$+30\degr47\arcmin00\arcsec (J2000)), where the dynamics of the gas is important.}
    \label{fig:veloHI+CO}
  \end{figure*}

  Figure \ref{fig:veloHI+CO} shows the CO velocity field, built in the same way, on top of the \ion{H}{I} peak velocity map. The CO velocity, shown within the contoured regions, is only determined where CO is detected above 4\,$\sigma$ (contoured in black). The absence of shifts between the \ion{H}{i} velocity (outside of contours) and the CO velocity  shows that CO and \ion{H}{I} are closely linked and are not separated by more than a few kilometers per second. This confirms that we can use the the atomic gas to trace the CO velocity.

  Subtracting the baseline yields the spectra from which a final data cube is created as well as the integrated intensity map. At the end of the data reduction process, we obtain a CO(2--1) cube at 12$\arcsec$ resolution with a 3$\arcsec$ pixel size and with velocity resolution of 2.6$\kms$ as well as the CO(2--1) integrated line intensity map (see Fig.~\ref{fig:ico0K}). We also produce a CO(2--1) cube and an integrated intensity map at 25$\arcsec$ resolution following the same procedure for a direct comparison with lower resolution observations such as the CO(1--0) data presented in Sect.~\ref{sec:co10}.

  \subsection{Noise map}
  
  \begin{figure}
    \centering
    \includegraphics[width=9cm]{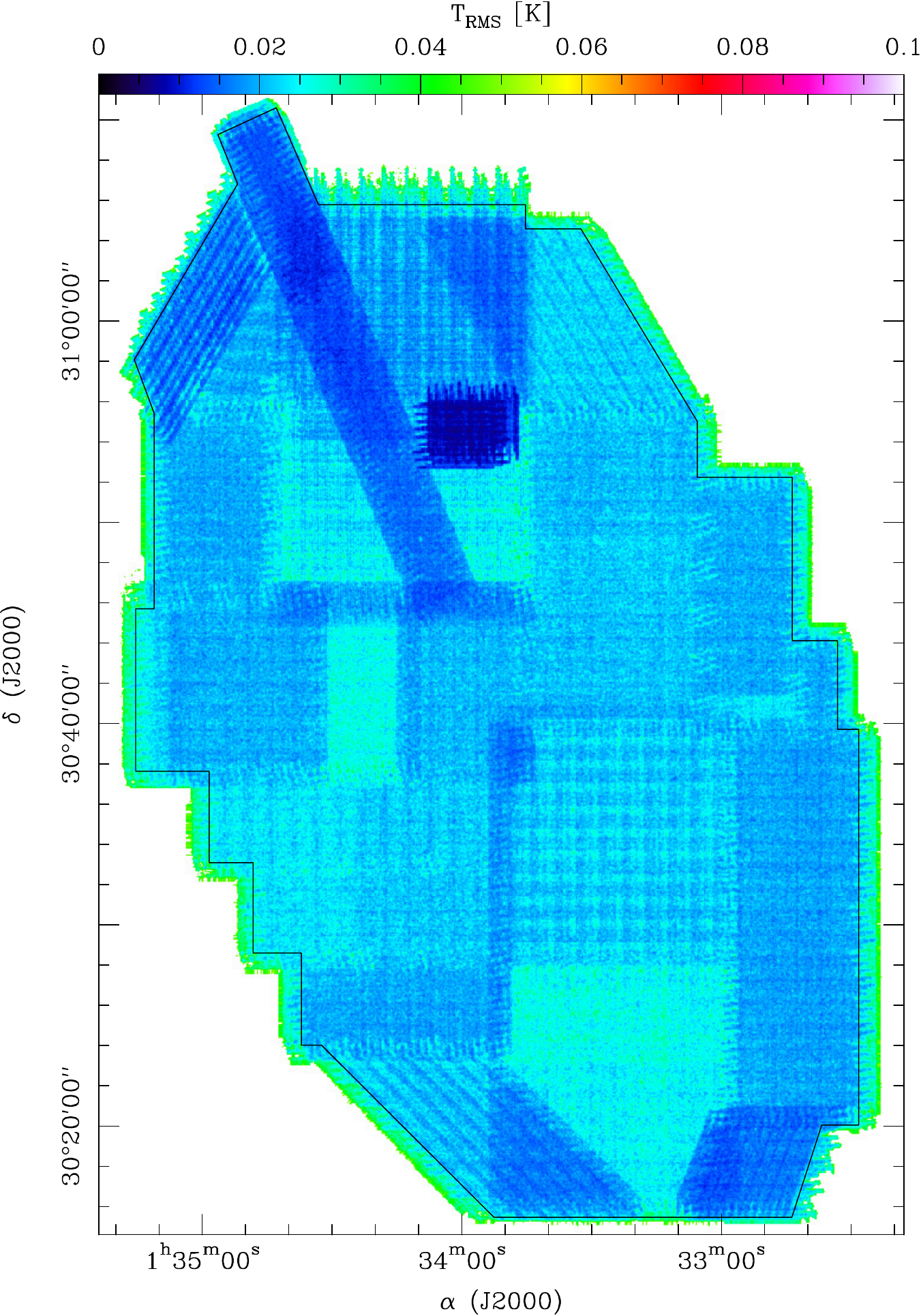}\\
    \caption{Noise map of the M~33 CO(2--1) data, in Kelvin per 2.6$\kms$ channel in antenna temperature. The average RMS noise per channel (inside the black contour) is 20.33~mK with a maximum of 50~mK at the map edge.}
    \label{fig:rms}
  \end{figure}

  Noise maps are computed using emission free channels of the spectra. For this we used the same mask based on \ion{H}{i} emission and the rotation curve as the one presented in Sect.~\ref{sec:reduction}. The fields with the lower noise correspond to a larger number of coverages and/or better weather conditions during the observations. The radial strip observed with higher sensitivity is the HerM33es strip partially observed in [\ion{C}{II}], [\ion{O}{I}] and [\ion{N}{II}] by Herschel. The variations we see in the noise level are emphasized by the color scale but only happen around a very low level. The mean RMS noise map level over the disk up to 7~kpc is 20.33~mK ($T_\mathrm{a}^*$) with a fairly homogeneous distribution which can be seen in Fig.~\ref{fig:rms}. The median of the noise distribution is 20.37~mK. 

  \subsection{CO(1--0) data}
  \label{sec:co10}
  The CO(1--0) transition at 115.271~GHz was observed during some poor but not terrible weather periods when CO(2--1) observing was not feasible, for a total of 40 hours. The Earth's atmosphere is less of a problem for observations of CO(1--0). The map is confined to the HerM33es strip, at a resolution of 0.81$\kms$, with an average system temperature of 300~K ($T_\mathrm{a}^*$).
  Data reduction was essentially the same as for the CO(2--1) - filtering of poor baselines and construction of a cube at $25\arcsec$ resolution. The CO(1--0) integrated intensity for maps at $25\arcsec$ were calculated using \ion{H}{i} intensities down to 0~K to define the line windows.

\section{Mass and distribution of the molecular gas}

  \subsection{Integrated intensity maps}

    \subsubsection{Determination of line windows}
    \label{sec:icocomput}

    Three methods were tested to compute the integrated CO(2--1) line intensity. The most basic is to use the rotation curve \citep{Corbelli1997} plus a pre-defined line window to determine the channels to be summed to make the integrated intensity map. The other two methods use the \ion{H}{i} line data \citep{Gratier2010a} as described in Sect.~\ref{sec:reduction} but testing two threshold signal levels : 2.6~K (1\,$\sigma$ noise level) and 0~K. We used the \ion{H}{i} cube to locate the \ion{H}{i} peak velocity for emission $\ge$ 10~K. The velocity limits are then given by the first channel below a certain threshold (2.6 or 0~K) on each side of the peak. We adopted the latter method (\ion{H}{i} line down to 0~K) for the reasons described below.

    Figure \ref{fig:veloHI} shows the velocity field of M~33 based on \ion{H}{i} peak temperature with the regions where the \ion{H}{i} does not reach 10~K ($\sim$4\,$\sigma$) indicated by contours. Within these contours, the rotation velocity is assumed to be that defined by the \citet{Corbelli1997} rotation curve as we consider that the velocity at a threshold below 10~K is less reliable than that of the rotation curve.
    Figure \ref{fig:deltaV} shows the width of the line window at all positions, determined with the 0~K \ion{H}{i} threshold where the peak \ion{H}{i} line temperature exceeds 10~K and chosen to be 60$\kms$ centered around the rotation curve velocity elsewhere. As shown in Fig.~\ref{fig:deltaV}, there are some regions where a 60$\kms$ width mask window would not be enough and many others where it would probably be too much.

    The line windows generated from the \ion{H}{i} emission, based on the assumption that CO emission could be present at all velocities at which \ion{H}{i} emission was detected, were sometimes very broad, reaching $\sim$100$\kms$ in a few regions actively forming stars (see Fig.~\ref{fig:deltaV}). Some of these regions are far from the center so that a simple radial decrease in the assumed linewidth could not be applied to the rotation curve. Using a line window encompassing all of the \ion{H}{i} velocities but based on the rotation curve results in large uncertainties in the integrated intensities. Windows larger than necessary make baseline subtraction much more error-prone, allowing baseline fluctuations within the line window. Thus, either very broad windows were used or some velocities at which \ion{H}{i} is detected were not included in the line windows if we base the windows on the analytical rotation curve.

    \citet{Gratier2010b} used this technique on NGC~6822 with a threshold \ion{H}{i} temperature of 10~K. CO emission in M~33 is much stronger than in NGC~6822 so the \ion{H}{i} threshold temperatures tested were lower: 2.6 and 0~K, where 2.6~K is the 1\,$\sigma$ noise level in the \ion{H}{i} cube. For most of the parts, the \ion{H}{i} and rotation velocities agree well as the differences are small (Fig.~\ref{fig:veloHI}). Double peaked spectra \citep[see e.g. spectra in the appendix of][]{Gratier2012} and regions where the \ion{H}{i} peak was significantly different from the analytical rotation curve were checked by eye to ensure that e.g. between two peaks the emission did not reach the 0~K level. The main potential advantage of using the more restrictive 2.6~K threshold is to frame the CO line more closely, reducing the noise in the integrated intensity map. However, the reduction in window width was small and we felt the potential risk of missing emission due to a dip between peaks or due to negative noise spikes outweighed the slight reduction in noise.

\begin{figure}
      \centering
      \includegraphics[width=9cm]{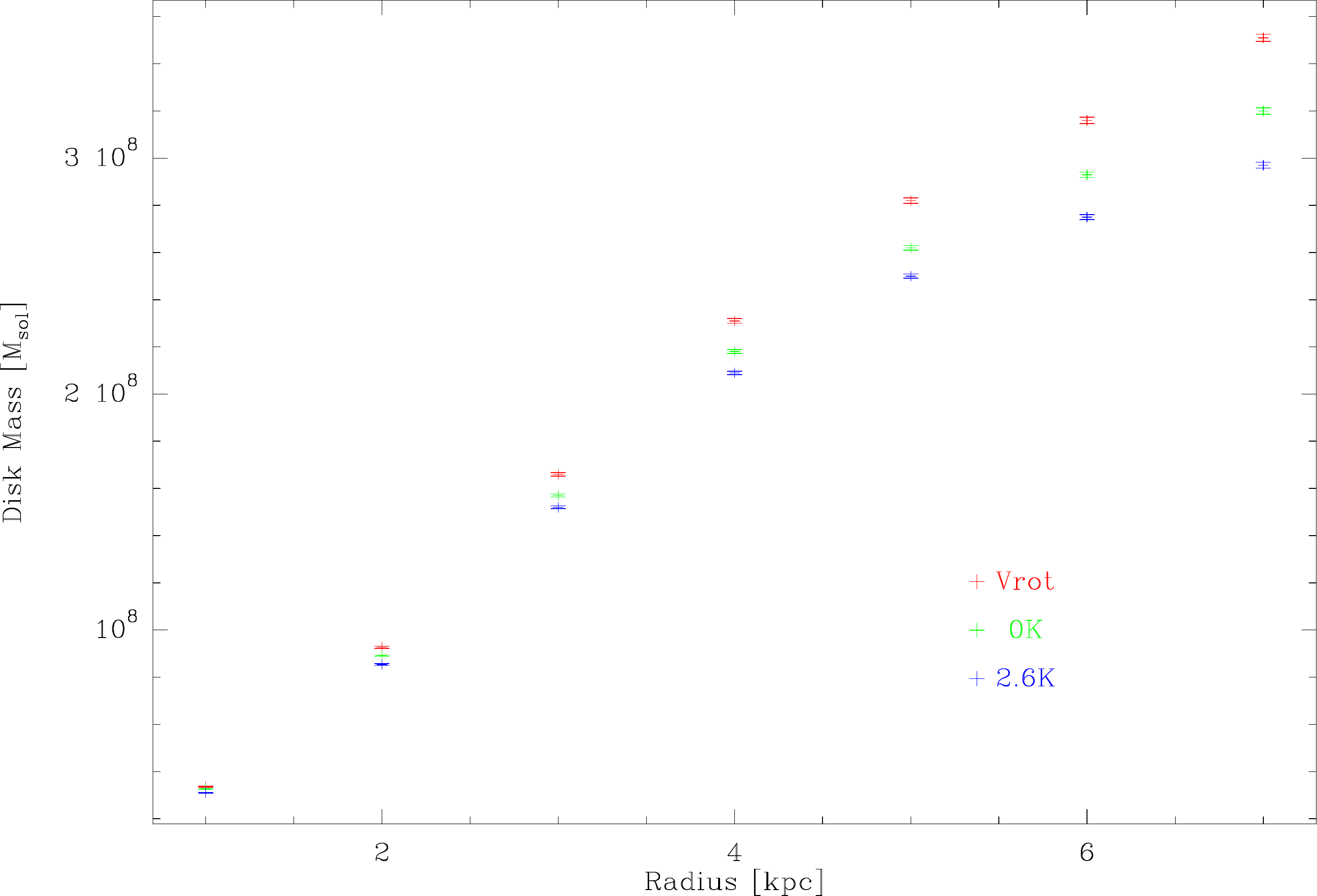}\\
      \caption{Intensities and the corresponding derived mass from the integrated CO(2--1) flux. This figure shows the total cumulative values contained in a given radius disk. The three symbols  correspond to different \ion{H}{i} windows used in the computation process. The conversion from CO luminosity to H$_2$ mass assumes a $\ratio = 4~\times~10^{20} \Xunit$ and a line ratio of 0.8 (see Sect.~\ref{sec:gasdistrib}). The error bars are based on the mean $\Delta I_\textrm{CO}$ over the disk as described in the last paragraph of Sect.~\ref{sec:icocomput}.}
      \label{fig:ico-lummass}
    \end{figure}

    Comparisons between the integrated intensity maps produced with only the rotation curve mask and with the \ion{H}{i} mask with thresholds of 0~K and 2.6~K (plus the rotation curve mask for the \ion{H}{i} holes) are presented in Fig.~\ref{fig:ico-lummass}. For \ion{H}{i} based masks, we can see that the 2.6~K-threshold mask apparently misses part of the CO emission as a difference can be seen with the 0~K curve. However, the highest intensities are given when computed with only the rotation curve mask. The telescope error beam picks up emission over a broad range in velocities so increasing the velocity range of the window results in an apparent increase in flux (see Sect.~\ref{sec:sidelobes} paragraphs 2 and 3).

\begin{figure}
      \centering
      \includegraphics[width= 9cm]{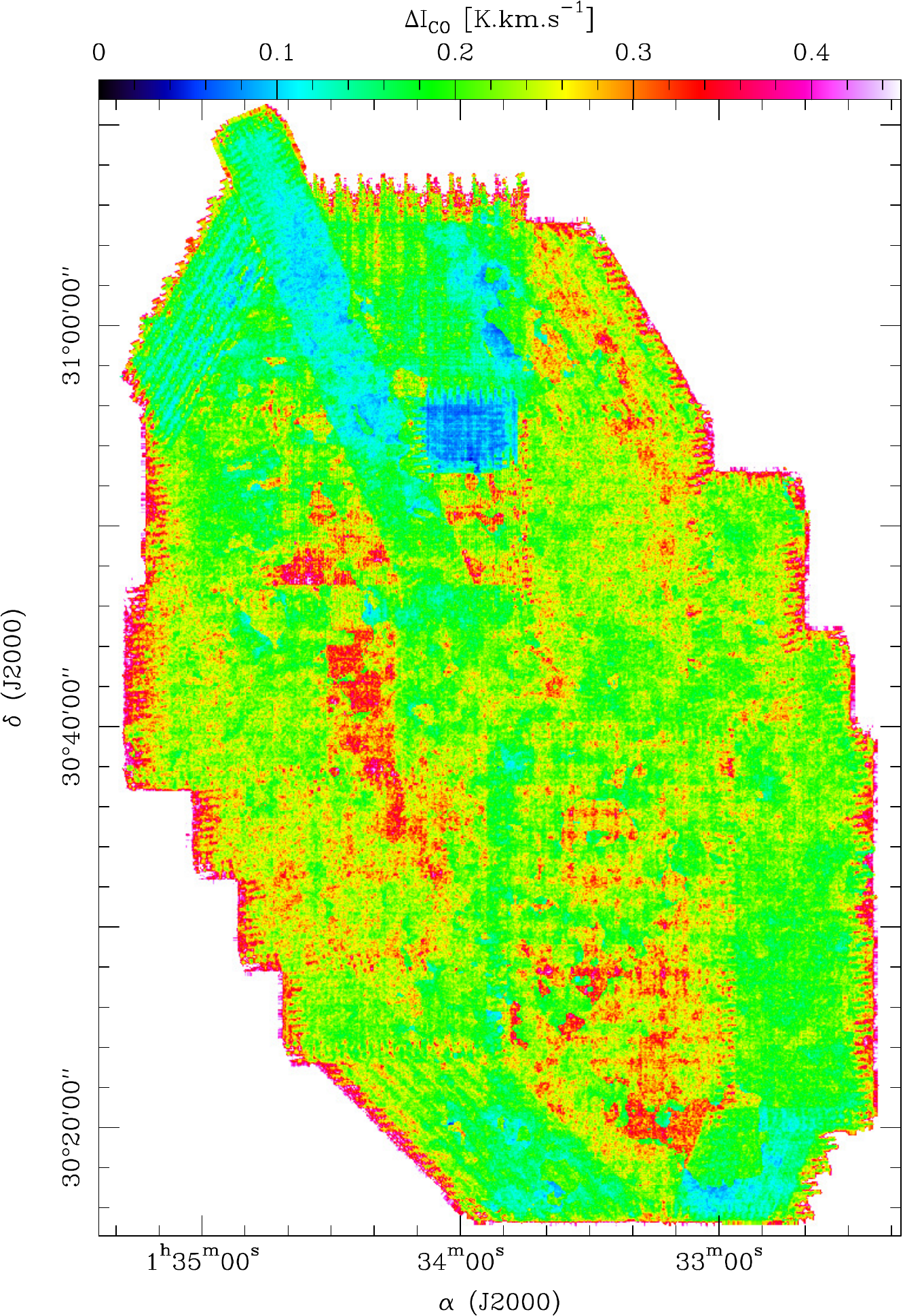}
      \caption{RMS noise of the integrated intensity map with \ion{H}{i}=0~K mask, in K$\kms$.}
      \label{fig:deltaico0K}
    \end{figure}

    The integrated intensity (moment zero) map produced with the \ion{H}{i} mask with a 0~K threshold is presented in Fig.~\ref{fig:ico0K} and is computed using main beam temperature. The uncertainty in integrated intensity $\Delta I_\mathrm{CO}$, Eq. 2, is shown in Fig.~\ref{fig:deltaico0K}, based on Figs.~\ref{fig:rms} and \ref{fig:deltaV}.
    \begin{equation}
        \Delta I_\mathrm{CO} = RMS \times \delta\nu_{\mathrm{ch}} \sqrt{N_{\mathrm{ch}}} = RMS \times \sqrt{\Delta\nu_{\mathrm{win}}.\delta\nu_{\mathrm{ch}}}
    \end{equation}
    where $RMS$, in Kelvin, is the noise value at a given pixel, $\delta\nu_{\mathrm{ch}}$ is the channel width (2.6$\kms$), $\Delta\nu_{\mathrm{win}}$ the window width in $\kms$ and $N_{\mathrm{ch}} = \frac{\Delta\nu_{\mathrm{win}}}{\delta\nu_{\mathrm{ch}}}$ the number of channels in the window. The mean $\Delta I_\mathrm{ CO}$ value on the disk is 0.22 K$\kms$ and 0.20~K$\kms$ for the regions where \ion{H}{i} is detected above 4\,$\sigma$.

    The error bars of Fig.~\ref{fig:ico-lummass} account for statistical uncertainties. The uncertainty is taken to be the average of $\Delta I_\mathrm{CO}$ (Fig.~\ref{fig:deltaico0K}, within the relevant annulus) divided by the square root of the number of lobes $\sqrt{N_\textrm{pix}/N_\textrm{pix/lobe}}$ and expressed as a mass of H$_2$.

    \subsubsection{Question of error beam pickup}
    \label{sec:sidelobes}

    The difference between the forward efficiency and the beam efficiency corresponds to power received in the error beams. M~33 is an extended source so some of the flux at any given position will come from error beam power towards other regions in M~33. The error beams of the IRAM 30 meter dish have been significantly reduced in strength since the \citet{Greve1998} publication due to major surface improvement in 2002. Calculations of the post-2002 error beam pattern can be found on the IRAM 30m web page in a report by Kramer et al. (2013)\footnote{\url{http://www.iram.es/IRAMES/mainWiki/CalibrationPapers?action=AttachFile&do=get&target=eb2013-v8.2.pdf}}. We compute an estimate of the emission from the error beams by adopting a three-Gaussians error beam structure from the report and calculating the pick-up out to the 6\% level of the broadest of the error beams (FWHM $\sim$800$\arcsec$). 

    Because the error beams pick up emission up to $\sim$800$\arcsec$ from the pointing center, the velocity of that emission does not necessarily correspond to that of the pointing center and thus to the line window.  Among the questions one might ask is whether the difference in flux between the different line windows used in Fig.~\ref{fig:ico-lummass} is due to emission picked up by the error beams at other velocities. This is important because more emission is picked up each time the window is broadened from the 2.6~K to 0~K threshold to a 60$\kms$ window. If this increase is not due to error beam pickup, then some CO emission must be present at velocities where the \ion{H}{i} emission is extremely weak (below 2.6~K) or not detected at all.

	  To estimate the possibility of a significant error beam pickup between our line masks we calculated the error beam emission cube. We measured the error beam flux as a function of the type of velocity windows we choose. With the analytical rotation curve parameters (a 60$\kms$ wide window), the error beam flux corresponds to about 2.5~$\times$~10$^7$~M$_\sun$ more than the 2.6~K \ion{H}{i} level windows and about 1.5~$\times$~10$^7$~M$_\sun$ more than the 0~K \ion{H}{i} level windows. This is very close to the differences observed in Fig.~\ref{fig:ico-lummass}, suggesting that the differences are indeed due to error beam pick-up. Therefore, it is not necessary to invoke CO emission at velocities outside the \ion{H}{i} range to explain the differences observed in Fig.~\ref{fig:ico-lummass}.

    Previous articles on M~33 have not addressed this issue so we do not attempt to subtract this emission in order to keep our maps comparable to those of earlier work. As we estimate the error beam pickup from the observed emission, which contains the error beam emission, this would be ideally accounted for by an iterative process. Such a detailed analysis is beyond the scope of this article. The uncertainty in flux due to error beam pickup is not constant because in regions with a high velocity gradient the error beam flux is more likely to fall outside the line window and be eliminated with the baseline subtraction.

    \subsubsection{\ion{H}{i}-poor regions}
    \label{sec:iconoHI}

  	A question that comes naturally is for the regions where the \ion{H}{i} emission is weak, not reaching 10~K ($\sim$4\,$\sigma$). These areas are marked by contours in Fig.~\ref{fig:ico0K} and the line window is set to 60$\kms$ (Fig.~\ref{fig:deltaV}). The \ion{H}{i}-poor regions of the disk represent about 7\% of the CO(2-1) coverage (1\% for the first kiloparsec). Is CO emission detected in these regions?
	
    In general, it is supposed that where \ion{H}{i} is not present, and gas density/pressure and metallicity are not high enough to cause complete conversion to H$_2$ (as can be the case in galactic nuclei), CO (H$_2$) is not expected to be present. This assumption can to some degree be tested as the error pattern should generate weak CO "emission" in the \ion{H}{i} holes. In this way, it is possible to estimate the amount of CO by summing the integrated intensity maps corresponding to these \ion{H}{i} holes. For these positions, the mask used is based on the analytical rotation curve of M~33. The total integrated signal over these areas corresponds to a value of 4.8~$\times$~10$^6$~M$_\sun$. We used the error beam cubes computed in the previous section to compare with this value. The integrated signal due to error beam on the very same regions is 5~$\times$~10$^6$~M$_\sun$ which we consider equivalent.

    The apparent CO emission we see in \ion{H}{i}-poor regions could be entirely due to error beam pickup, such that we have no evidence for CO detection where the peak \ion{H}{i} line temperature does not reach 10~K.

  \subsection{CO($\frac{2-1}{1-0}$) line ratio}
  \label{sec:lineratio}
    
  \begin{figure}
    \centering
      \includegraphics[width=8cm]{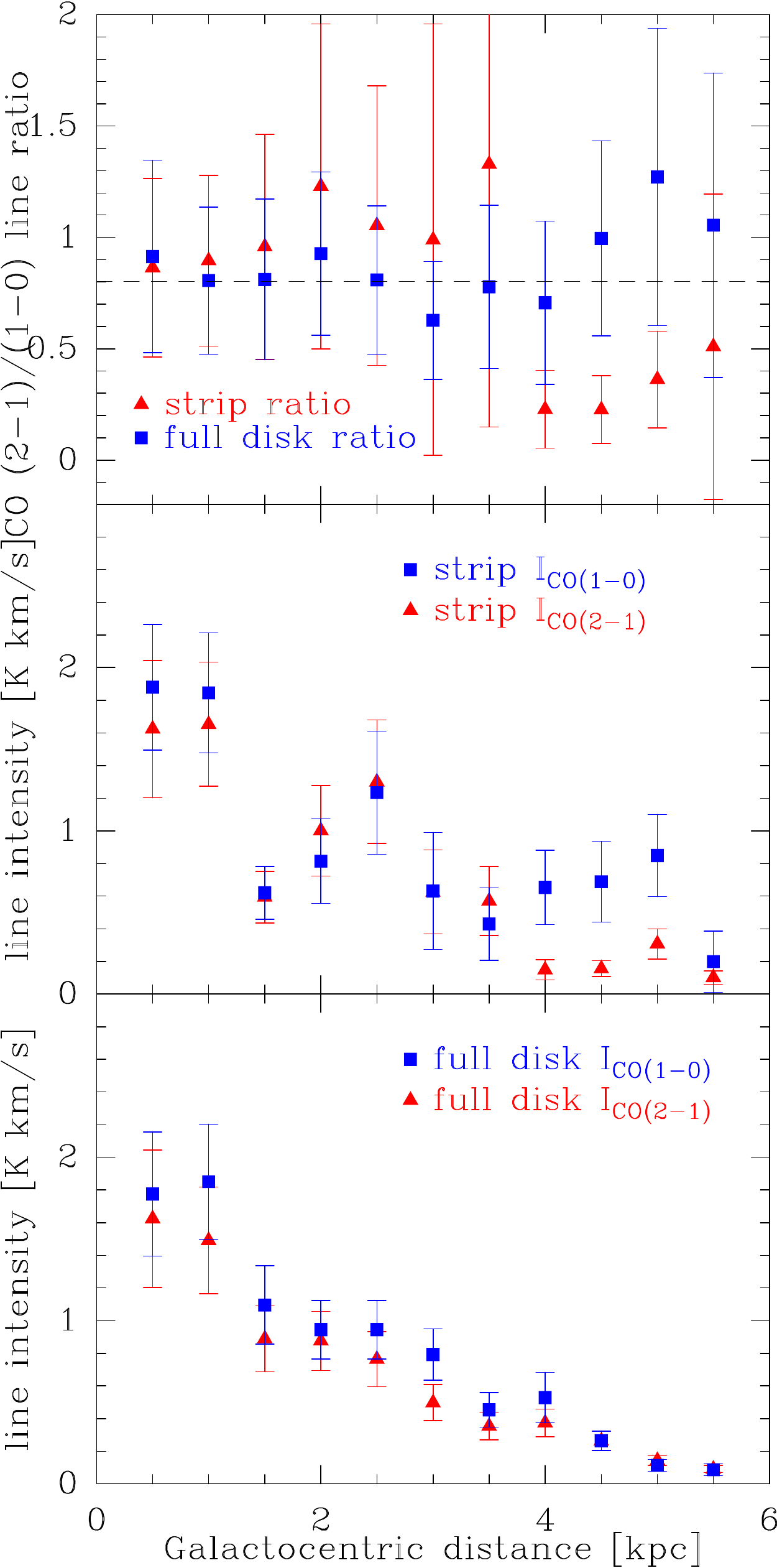}
      \caption{\emph{Top} Line ratios along the HerM33es strip (triangles) and for annuli of 0.5~kpc width (stars). The dashed line represents the 0.8 value we assume for the ratio in this paper. \emph{Middle} Radial evolution of the CO(2--1) and CO(1--0) line intensity along the strip. \emph{Bottom} Radial evolution of the CO(2--1) and CO(1--0) line intensity for the full disk. For all the figures, strip CO(1--0) data is described in Sect.~\ref{sec:co10} and disk CO(1--0) data is taken from \citet{Rosolowsky2007}. Error bars are derived from statistical and calibration uncertainties and are described in the 2\textsuperscript{nd} paragraph of Sect.~\ref{sec:lineratio}.}
      \label{fig:lineratios}
  \end{figure}

  Figure \ref{fig:lineratios} shows CO($\frac{2-1}{1-0}$) line ratio as a function of radius both for the whole disk and for the HerM33es strip. The CO(2--1) data used here are from the cube calculated at an angular resolution of $25\arcsec$ and the CO(1--0) data along the strip were presented in Sect.~\ref{sec:co10}, are at a resolution of $25\arcsec$. The CO(1--0) values for the entire disk are however derived from the masses given by \citet{Rosolowsky2007} in Fig.~7 (although their coverage is not full above 4~kpc). Their data were divided by 4.3 to convert them from M$_\sun$pc$^{-2}$ to cm$^{-2}$K$\kms$ accounting for Helium and corrected from an inclination of $52\degr$ to the $56\degr$ we assume. The line ratios are then calculated for each radial bin for the whole map and for the strip. The middle plot shows that although the behavior of the line ratio of the strip is not like that of the whole disk, the line intensities of the two lines follow each other very well, showing that the line ratio varies over the disk and not just with radius. The bottom plot of Fig.~\ref{fig:lineratios} shows the decrease of the line intensity with the distance from the center.

  We estimate the uncertainty in the calibration of the data to be 15\% for the CO(2--1), 10\% for the CO(1--0) IRAM data and 15\% for the \citet{Rosolowsky2007} CO(1--0) data. Statistical variations have also been taken into account by including the RMS scatter in the integrated intensity map. Uncertainties are derived by dividing this scatter by the square root of the number of lobes in the covered area $\sqrt{N_\textrm{pix}/N_\textrm{pix/lobe}}$. Statistical uncertainties for full disk CO(1--0) data are taken from Fig.~7 of \citet{Rosolowsky2007}. The statistical uncertainties on these values are generally larger than in Fig.~\ref{fig:ico-lummass} because they account for the whole scatter in $I_\mathrm{CO}$ and not $\Delta I_\mathrm{CO}$. These values are thus the upper limit of the uncertainty (the larger the distribution of values is over a bin, the larger the uncertainty would be).

  While the average line intensities clearly decrease with radius, the CO($\frac{2-1}{1-0}$) ratio does not vary in a regular way. We assume a constant line ratio of 0.8, consistent with Fig.~\ref{fig:lineratios}, but higher than the 0.73 used by \citet{Gratier2010a}. In large spirals, the CO($\frac{2-1}{1-0}$) line ratio decreases radially away from the center \citep{Sawada,Braine1997}. The lack of a decrease may be due to the lower metallicity, and presumably lower average optical depth of the CO lines, in M~33.

	\subsection{Molecular gas distribution}
  \label{sec:gasdistrib}
  As with most work on CO observations, we will use the CO emission as a proxy for the H$_2$ column density, assuming a constant $\ratioo$ factor such that $N_\mathrm{H_2} = I_\mathrm{CO(2-1)}~\times~\ratiot$. The H$_2$ mass is then:
  \begin{equation}
    M_\mathrm{H_2} = I_\mathrm{CO(2-1)} \times \left(\frac{I_\mathrm{CO(2-1)}}{I_\mathrm{CO(1-0)}}\right)^{-1}  X_\mathrm{CO} \frac{2\mathrm{m}_\mathrm{p}}{f_\mathrm{mol}} \Omega \mathrm{D}^2
  \end{equation}
  where $I_\mathrm{CO(2-1)}$ is the CO(2--1) intensity on the main beam scale in K~km~s$^{-1}$, $\frac{I_\mathrm{CO(2-1)}}{I_\mathrm{CO(1-0)}}$ is the line ratio studied in Sect.~\ref{sec:lineratio} and taken equal to 0.8 throughout the disk. We take the CO to H$_2$ conversion factor to be $X_\mathrm{CO} = \frac{N_\mathrm{H_2}}{I_\mathrm{CO(1-0)}} = 4~\times~10^{20} \Xunit$, twice the Milky Way value to be consistent with the previous work of \citet{Gratier2010a}, assuming an inverse relation between $X_\mathrm{CO}$ and the metallicity \citep{Wilson1995}, and based on the Far-IR dust emission\citep{Braine2010}. $f_\mathrm{mol}$ accounts for the mass of the Helium in the molecular gas, with a correction of 37$\%$, and $\mathrm{m}_\mathrm{p}$ is the mass of the proton. D $=$ 840~kpc is the distance to M~33 and $\Omega$ is the beamsize in steradians.

  \begin{figure}
      \centering
      \includegraphics[width=9cm]{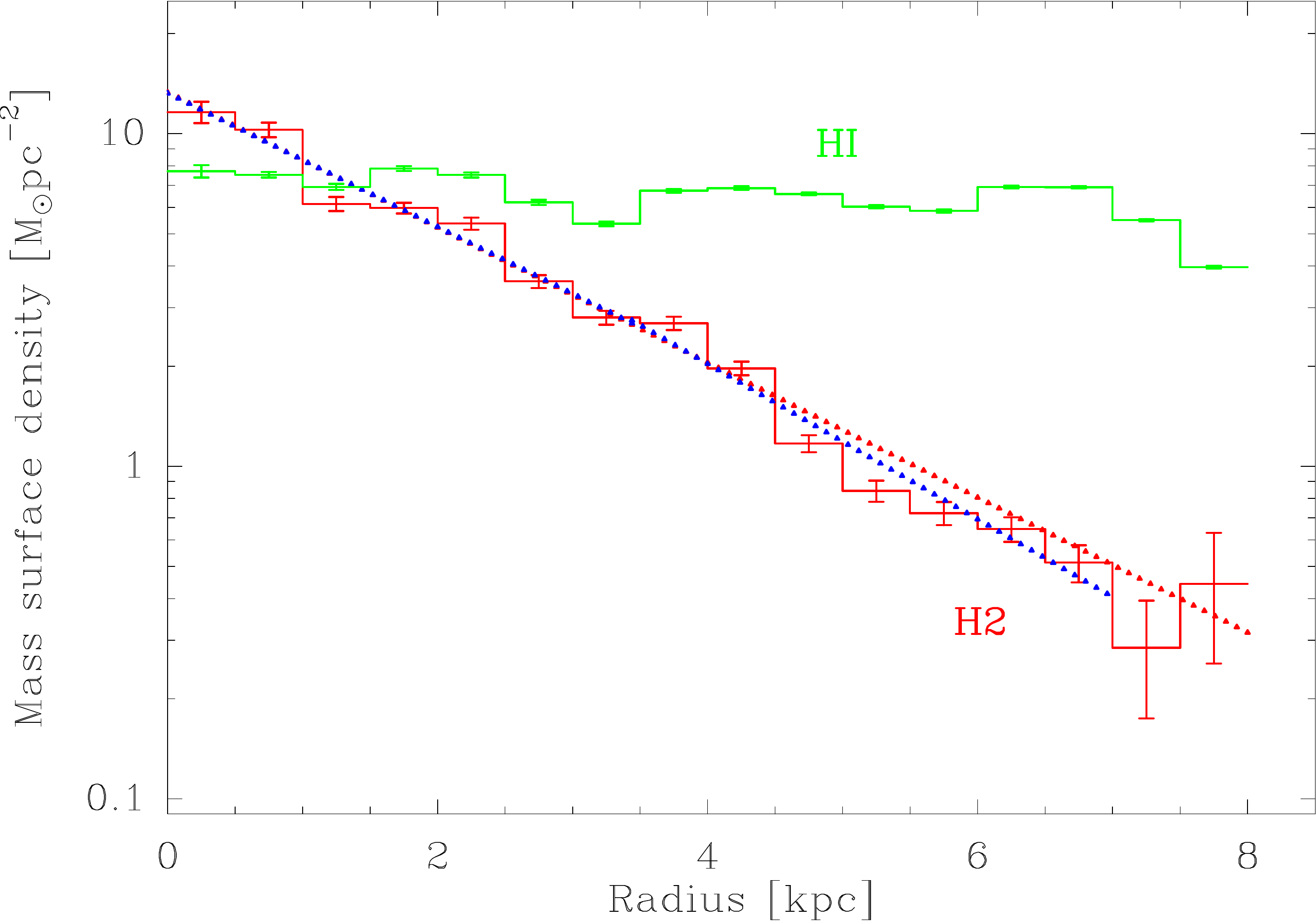}\\
      \caption{Radial distribution of the CO-derived H$_2$ (red) and \ion{H}{i} (green) mass surface density in M$_\sun$pc$^{-2}$. The mass surface density is corrected for inclination and includes Helium. The red dots correspond to a exponential fit to the first 7~kpc, yielding a exponential length of 2.1~kpc. The blue dots represent two partial fits: one for the inner galaxy (out to 3.5~kpc), and one for the outer galaxy (3.5~kpc < R < 7.0~kpc). Statistical error bars are derived from RMS scatter over the rings, as described in the 2\textsuperscript{nd} paragraph of Sect.~\ref{sec:gasdistrib}.}
      \label{fig:ico-exp}
  \end{figure}
  
  The azimuthally averaged radial distribution of the CO emission, and thus presumably H$_2$ mass, is shown in Fig.~\ref{fig:ico-lummass}. 
  Fitting an exponential disk (solid line) yields a disk scale length of 2.1$\pm$0.1~kpc when fitting the whole disk to 7~kpc (see Fig.~\ref{fig:ico-exp}). Fits can be done on the halves of the disk up to 3.5~kpc and beyond. This gives scale lengths of 2.2$\pm$0.3~kpc (0-3.5~kpc) and 1.9$\pm$0.2~kpc (3.5-7~kpc), similar to the results shown in \citet{Gratier2010a}. Error bars correspond only to statistical variations, to be comparable with \ion{H}{I} data, and are equal to the RMS scatter in $I_\mathrm{CO}$ divided by the square root of the number of beams in the covered area $\sqrt{N_\textrm{pix}/N_\textrm{pix/lobe}}$.

  While the H$_2$ surface density, assuming we have chosen an appropriate $\ratioo$ conversion factor, is slightly higher than that of the atomic gas in the center, the \ion{H}{i} is dominant beyond the inner kiloparsec.

\section{Relationship between atomic and molecular gas}
\label{sec:recenter}
  
  How closely linked are the atomic and molecular components? If molecular clouds are seen as dense clumps embedded in a much warmer and more diffuse neutral atomic medium, then their velocities are not necessarily linked. On the other hand, if the H$_2$ forms quiescently from the denser atomic clouds, then one would expect that the dispersion between the two components would be very small. The interstellar medium of M~33 is dominated by the atomic component virtually throughout, unlike e.g. M~51, such that the H$_2$ in M~33 forms from the \ion{H}{i} rather than the \ion{H}{i} forming from photo-dissociated H$_2$.

  Molecular gas is not always found where the \ion{H}{i} column density is high -- there are high N(\ion{H}{i}) regions without CO emission just like molecular clouds are sometimes observed in regions of moderate \ion{H}{i} column density \citep[e.g. "Lonely cloud",][]{Gardan2007}. This suggests that other factors play a role in provoking the conversion of \ion{H}{i} into H$_2$. Various schemes have been suggested to explain large-scale atomic-to-molecular gas ratios \citep{Blitz2006,Gnedin2009,Krumholz2008} in galaxies but not the formation of individual GMCs. It has been suggested that "colliding flows" \citep{Elmegreen1993,Audit2005,Clark2012} or "colliding clouds" (Motte et al., 2014, submitted) may provide the compression required to create H$_2$ from \ion{H}{i}.  Presumably, this process would increase linewidths proportionally to the shock velocity and increase the dispersion between the \ion{H}{i} and H$_2$ velocities.

  In order to estimate the velocity dispersion between the \ion{H}{i} and CO, we have subtracted the \ion{H}{i} velocity, as measured by the velocity at the line peak, from the CO spectra. The new cube of CO(2--1) spectra "recentered" to the \ion{H}{i} velocity can be used to stack spectra. In this way, we directly obtain the \ion{H}{i}-CO velocity dispersion, once a typical CO linewidth has been established. This technique, used in \citet{Schruba2011} and \citet{Caldu2013} is also very useful to see large scale radial variations or reveal low-level emission through coherent stacking over large areas.  

  In addition to recentering the CO line with the \ion{H}{I} velocity, we applied this technique to the CO cube using the 4\,$\sigma$ detections of CO to determine velocities and recentering the CO with the CO.  Similarly, the \ion{H}{i} cube was recentered using the velocity of the \ion{H}{i} peak temperature.

  \subsection{Recentered cubes : method}

  For each line of sight (each spectrum in our regularly gridded data cube), we can associate a velocity based on the \ion{H}{i} (or on the analytical rotation curve -- see below) and the CO spectrum at this position can be shifted from this velocity to a reference velocity set to be 0$\kms$ by redefining the velocity axis.

  The velocity used to recenter the spectra can be defined in different ways: ($i$) with the analytical rotation curve \citep[Eq. 4 of][]{Corbelli1997}) given for each point in the disk, ($ii$) the peak \ion{H}{i} channel velocity (computed and used in Sects.~\ref{sec:reduction} and \ref{sec:icocomput}) completed with the analytical rotation curve for regions where there is no \ion{H}{i} over 10~K (see Fig.~\ref{fig:veloHI}) or ($iii$) the first moment of the \ion{H}{i} emission calculated in a window within 30$\kms$ of the rotation curve as following: 
  \begin{equation}
    \langle V \rangle= \frac{\Sigma_{\mathrm{channels}} Tv\mathrm{d}v}{\Sigma_{\mathrm{channels}} T\mathrm{d}v}
  \end{equation}
  where $T$ is the temperature of the channel and $\nu$ the frequency associated with the channel.
  Over 90\% of the disk of M~33 has \ion{H}{i} spectra with S/N $> 4$ so that the velocity of the peak temperature is well defined. The CO emission covers a much lower fraction of the disk as the vast majority of the lines of sight do not show CO above a 3\,$\sigma$ level.

  Once all the spectra have been recentered, they can be averaged (i.e. stacked). If the CO emission systematically follows the \ion{H}{i}, then we expect to find the CO peak of the stacked spectra at zero velocity. Any systematic difference between \ion{H}{i} and CO would create a velocity difference which, while invisible in the individual spectra, might appear in the stacked spectra. The width of the stacked CO spectra comes from the sum of the intrinsic width of the CO spectra, the intrinsic dispersion between the atomic and molecular components, the presence of multi-peaked CO, asymmetric profiles and the error in estimating the \ion{H}{i} velocity due to noise, in addition to the small broadening due to the finite channel widths.

\begin{figure}
    \centering
    \includegraphics[width= 9cm]{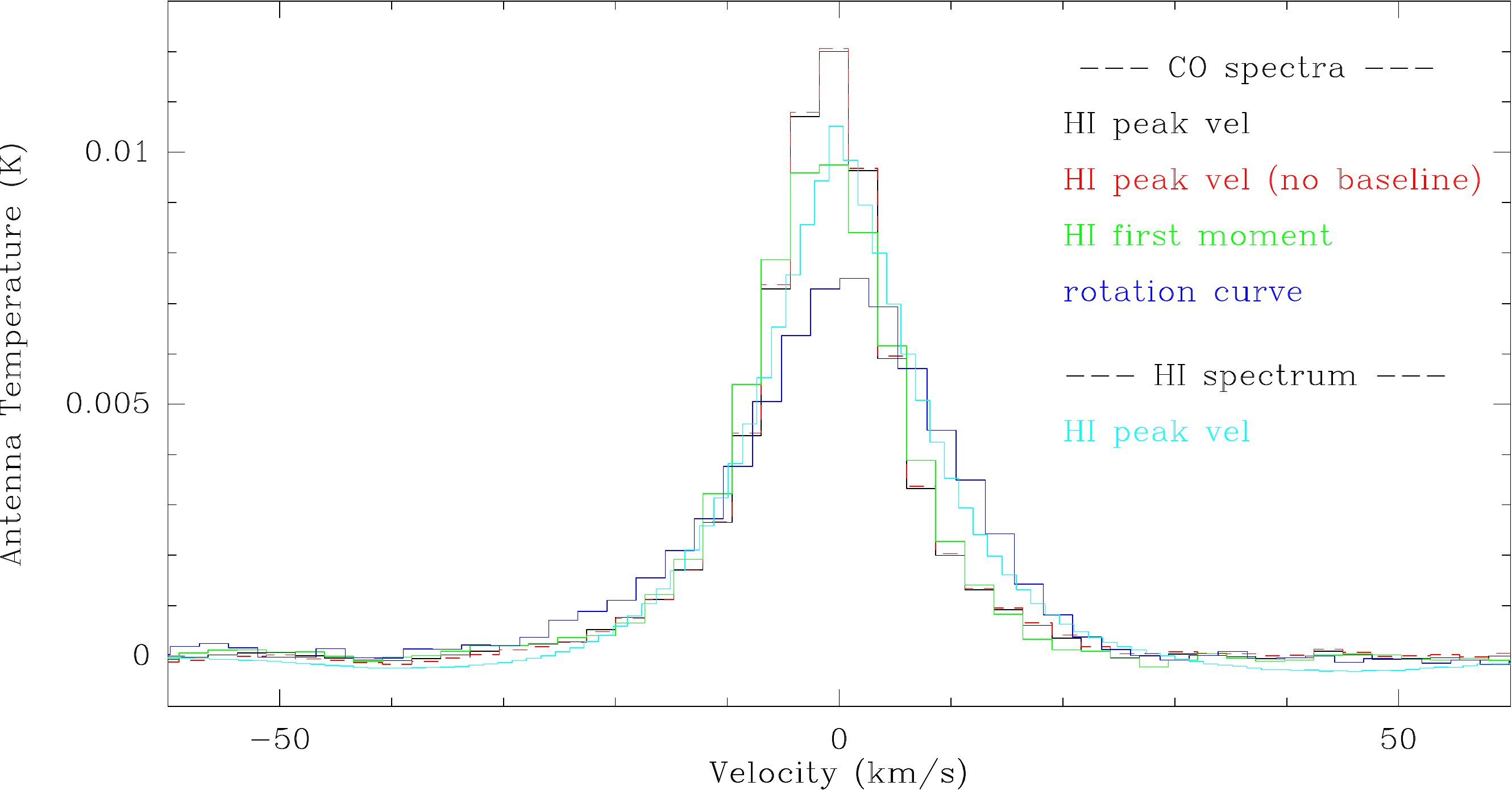}
    \caption{Averaged "recentered" spectra corresponding to the whole disk. Spectra are calculated from \ion{H}{i} spectra recentered on the \ion{H}{i} peak velocity (light blue) and from the CO cube recentered with the \ion{H}{i} peak velocity (black and red), the \ion{H}{i} emission moment velocity (green) and the analytical rotation curve velocity (dark blue). The \ion{H}{i} spectrum is divided by 2500 to be compared with the CO spectra.}
    \label{fig:recenterall}
  \end{figure}

  Figure \ref{fig:recenterall} shows the stacked spectra corresponding to the full disk coverage of M~33. This means that the {\it entire} disk of M~33 is included in these stacked spectra. Each spectrum was computed via a different centering method (see above) applied to the \ion{H}{i} and CO spectra: \ion{H}{i} peak line temperature, \ion{H}{i} first moment, and analytical rotation curve. The "\ion{H}{i} peak vel (no baseline)" spectrum comes from a baseline-free CO cube in order to make sure that the subtraction of a 3\textsuperscript{rd} order baseline does not affect the line wings. Because baseline variations are frequency dependent, the effect should statistically cancel each other out when stacked. In the end, the average spectra for the two "\ion{H}{i} peak vel" CO data cubes are the same (black and red in Fig.~\ref{fig:recenterall}), showing that our reduction process does not introduce any "false" signal in our data cube nor filter out anything potentially real.

  The \ion{H}{I} profiles are wider than those of the CO. This can be explained by the fact that the \ion{H}{I} clouds are bigger than the corresponding CO ones and that the velocity dispersion is higher.
  
  \subsection{CO-\ion{H}{i} : velocity dispersion and kinematics}

  This powerful technique also has the advantage of showing the CO velocity dispersion compared to the \ion{H}{i}. This parameter is related to the width of our stacked spectra. For example, a CO spectrum with a peak at a velocity different from the \ion{H}{i} velocity used during the centering process will broaden the stacked line. 

\begin{figure*}
    \centering
    \includegraphics[width=6cm]{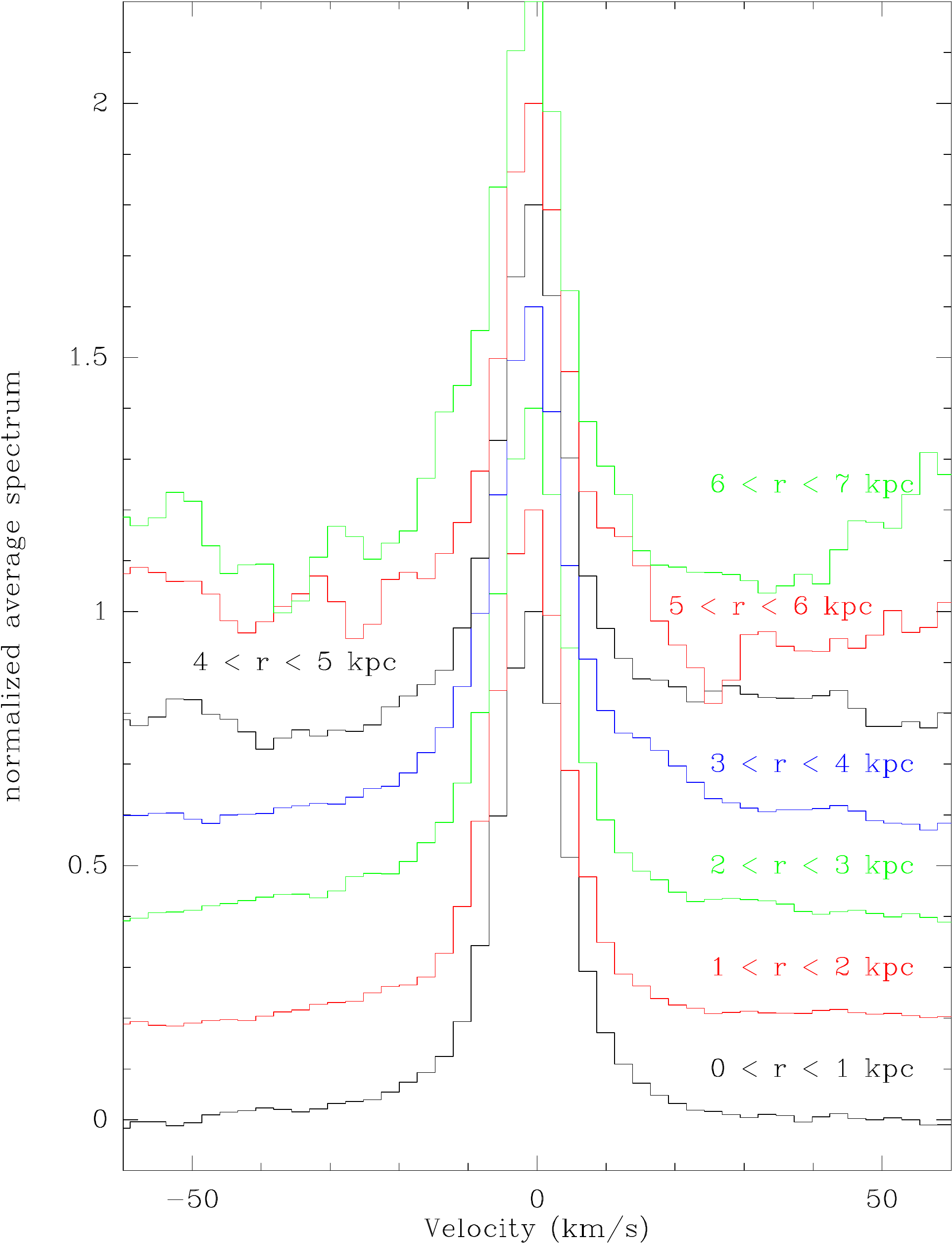}
    \includegraphics[width=6cm]{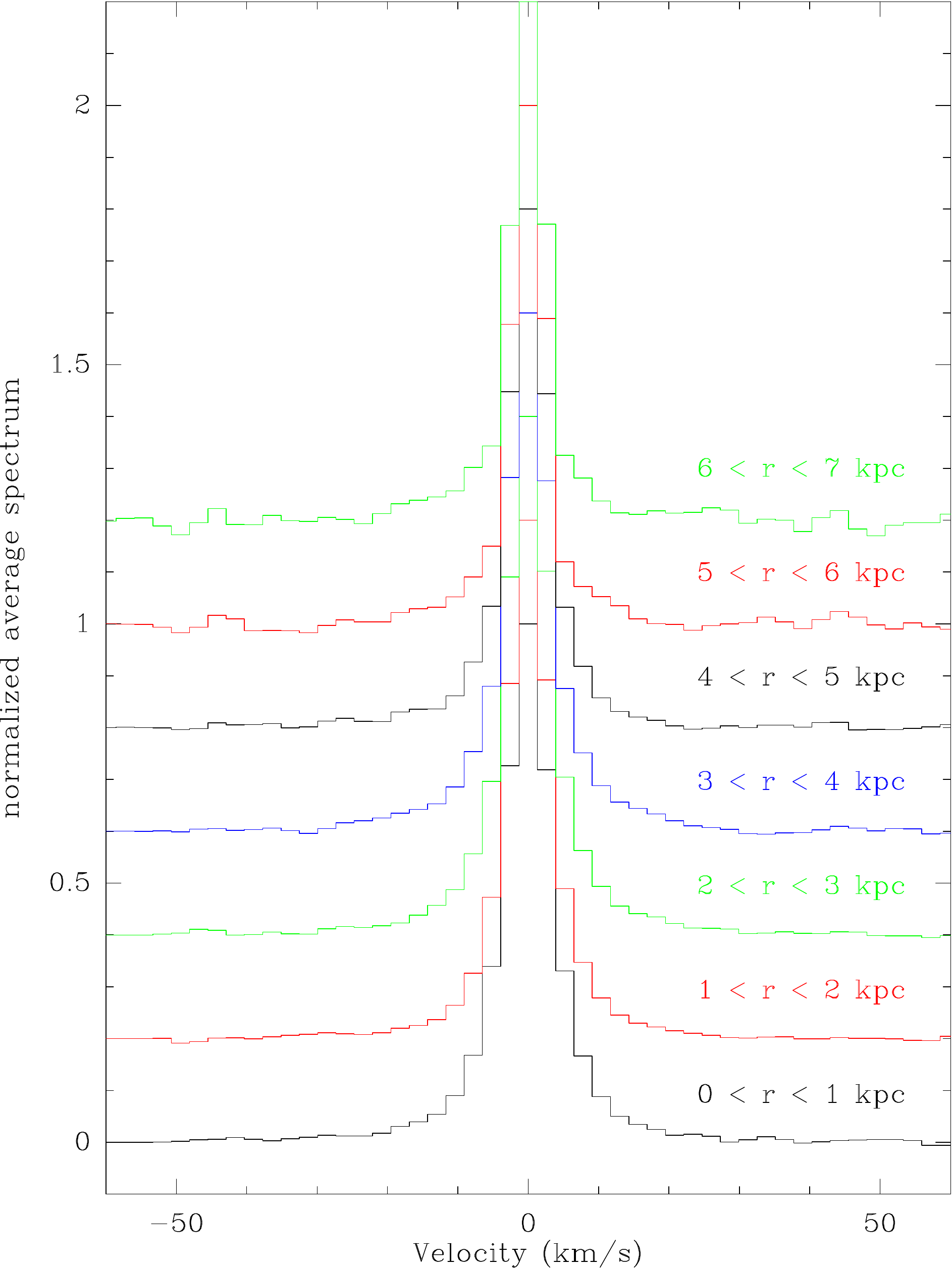}
    \includegraphics[width=6cm]{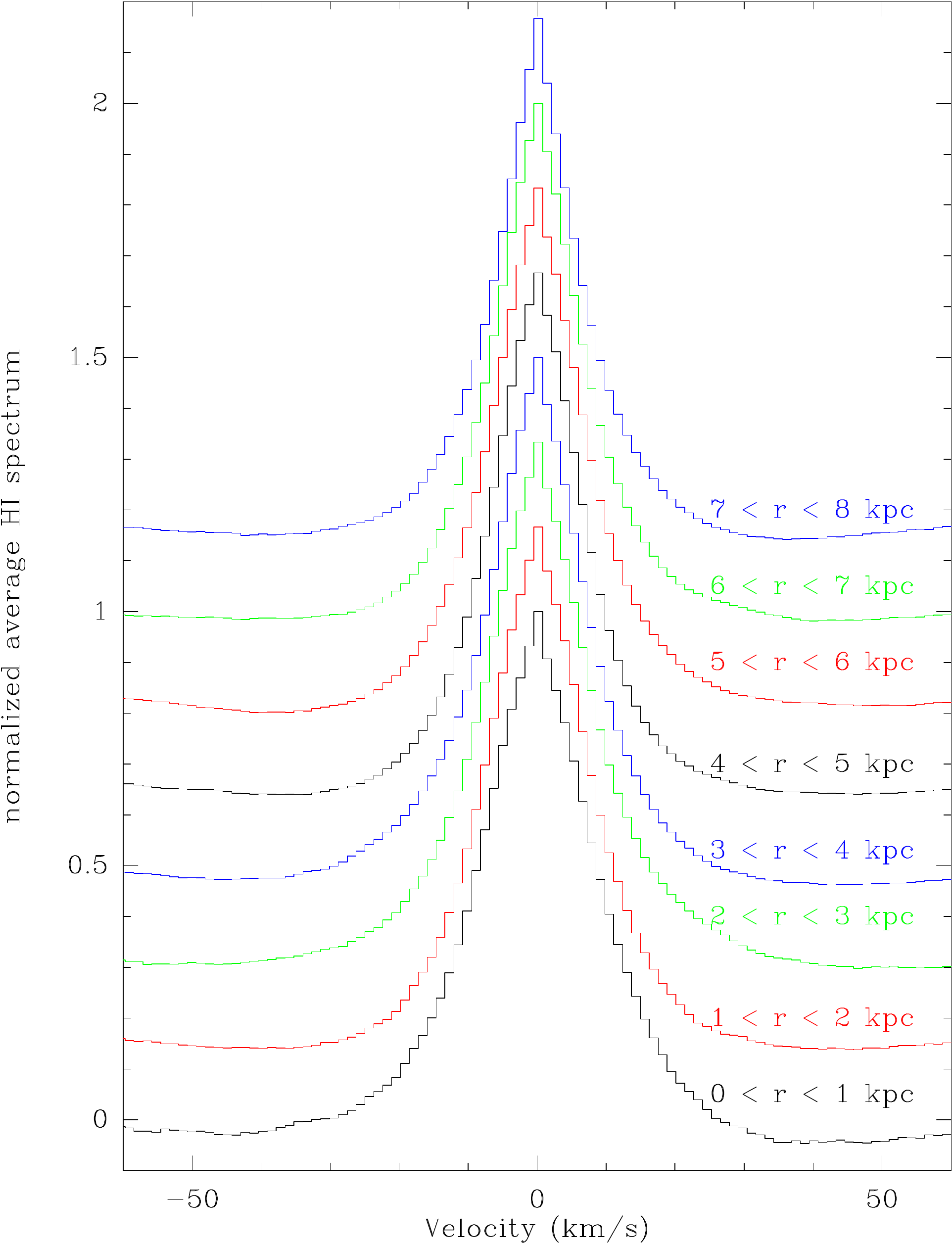}
    \caption{\emph{Left figure.} Average CO line profiles for each 1~kpc ring computed from the \ion{H}{i} peak velocity recentered CO(2--1) cube. \emph{Middle figure.} Average CO line profiles for each 1~kpc ring computed from the CO peak velocity recentered CO(2--1) cube. \emph{Right figure.} Average \ion{H}{i} line profiles for each 1~kpc ring computed from the recentered \ion{H}{i} cube. Each spectrum is normalized to unity and separated by adding 0.2.}
    \label{fig:line-recenter}
  \end{figure*}

  Gaussian profiles are fit to the central channels of each line ($\approx\pm 13\kms$ around the line center) down to the $\sim$6\% level. Table \ref{tab:linefits} shows the parameters of these fits : $\Delta V$ is the full width at half maximum and $V_\mathrm{max}$ the velocity shift of the peak of the line, compared to 0 and is computed as $V_\textrm{CO peak} - V_\textrm{\ion{H}{I} peak}$. Even though \ion{H}{i} line profiles are clearly not Gaussian (as it can also be seen on the right of Fig.~\ref{fig:line-recenter}), the \ion{H}{i} linewidth measurements can be reliable.
  The uncertainties of the fits are given by the MFIT algorithm of GILDAS.

  \begin{table}
    \caption{Gaussian fit parameters for the stacked CO(2--1) and \ion{H}{i} lines}
    \label{tab:linefits}
    \centering
    \begin{tabular}{l l | c c}
      \hline
      $V_{\mathrm{center}}$ & area & $\Delta V$ & $V_\mathrm{max}$ \\
      & & [$\kms$] & [$\kms$] \\
      \hline
      \multicolumn{4}{c}{CO(2--1) line} \\
      \hline
      $V_{\mathrm{\ion{H}{i} peak}}$ & 0-1~kpc & 12.4$\pm$0.4 & -0.2$\pm$0.6 \\
       & 1-2~kpc & 12.4$\pm$0.4 & -0.6$\pm$0.6 \\
       & 2-3~kpc & 13.0$\pm$0.5 & -0.5$\pm$0.6 \\
       & 3-4~kpc & 13.2$\pm$0.6 & -0.5$\pm$0.6 \\
       & 4-5~kpc & 11.7$\pm$0.4 & -0.1$\pm$0.6 \\
       & 5-6~kpc & 11.6$\pm$0.6 & -0.2$\pm$0.6 \\
       & 6-7~kpc & 12.1$\pm$0.9  & -1.0$\pm$0.6 \\
       & full disk & 12.5$\pm$0.4 & -0.4$\pm$0.6 \\
       & no baseline & 12.5$\pm$0.5 & -0.4$\pm$0.6 \\
      $V_{\mathrm{CO peak}}$ & 0-1~kpc & 8.2$\pm$0.3 & -0.03$\pm$0.09 \\
       & 1-2~kpc & 7.6$\pm$0.3 & -0.04$\pm$0.09 \\
       & 2-3~kpc & 7.8$\pm$0.3 & 0.03$\pm$0.10 \\
       & 3-4~kpc & 7.6$\pm$0.3 & -0.02$\pm$0.10 \\
       & 4-5~kpc & 7.0$\pm$0.3 & -0.01$\pm$0.10 \\
       & 5-6~kpc & 6.0$\pm$0.2 & 0.00$\pm$0.06 \\
       & 6-7~kpc & 5.9$\pm$0.2 & -0.01$\pm$0.06 \\
       & 7-8~kpc & 5.0$\pm$0.2 & -0.13$\pm$0.07 \\
       & full disk & 7.1$\pm$0.3 & 0.02$\pm$0.1 \\
      $V_{\mathrm{\ion{H}{i} mom}}$  & full disk & 15.1$\pm$0.1 & -0.8$\pm$0.6 \\
      $V_{\mathrm{rot}}$     & full disk & 19.8$\pm$0.3 & 0.7$\pm$0.1 \\
      \hline
      \multicolumn{4}{c}{\ion{H}{i} line} \\
      \hline
      $V_{\mathrm{\ion{H}{i} peak}}$ & full disk & 14.8$\pm$0.5 & -0.02$\pm$0.64 \\
      \hline
    \end{tabular}
  \end{table}

  Figure \ref{fig:recenterall} and Table \ref{tab:linefits} clearly show that CO linewidths are smallest when using the peak \ion{H}{i} velocity for the recentering. Compared to the \ion{H}{i} first moment velocity, this shows that H$_2$ is more likely to form at the velocity of the \ion{H}{i} line peak, presumably reflecting the velocity of the highest volume density material rather than at the most representative velocity (the first moment velocity). This phenomenon is seen when examining the cloud catalog for M~33 given by \citet{Gratier2012} where the majority of single CO peaks are associated with the stronger \ion{H}{I} peak when there are multiple peaks (e.g. clouds 256, 260, 282, 290) or when two \ion{H}{I} peaks match with two CO peaks (e.g. clouds 52, 45, 218). Nonetheless, in a few cases, a single CO peak can be located at the weaker \ion{H}{I} peak (e.g. clouds 209, 237, 225) or more rarely between two \ion{H}{I} peaks (e.g. cloud 163).

  The difference in linewidths (12.5$\pm$0.4$\kms$ versus 15.1$\pm$0.1$\kms$) is beyond what might result from the effect of noise in determining the \ion{H}{i} velocity. Recentering with respect to the analytical rotation curve, which is symmetric and determined by fitting the velocity field over the disk, yields a considerably broader line. This is not very surprising as approaching and receding halves of galaxies often do not show identical rotation curves; the comparisons here show that the small-scale wiggles in the rotation curve are followed by both the atomic and molecular components. The width of this "recentered" CO line can be seen as representing the sum of the average CO cloud velocity width and the cloud-cloud dispersion in an axisymmetric potential with no perturbations, although this may not be realistic.  
  
  It is also interesting to look at how this CO-\ion{H}{i} velocity dispersion evolves with the radius as shown in Figs.~\ref{fig:line-recenter} for CO and \ion{H}{i}. We have masked successively the emission outside of concentric rings of 1~kpc width and then summed all the individual spectra in the rings. In these figures, the intensities (normalized to unity) for each ring have been plotted, changing colors and adding 0.2 between successive rings. The presence of higher noise in the larger rings can be explained by the fact that there is less CO emission in regions further away from the center. The parameters of the Gaussian fits parameters are given in Table \ref{tab:linefits} for the stacked CO spectra.

\begin{figure}
    \centering
    \includegraphics[width=8cm]{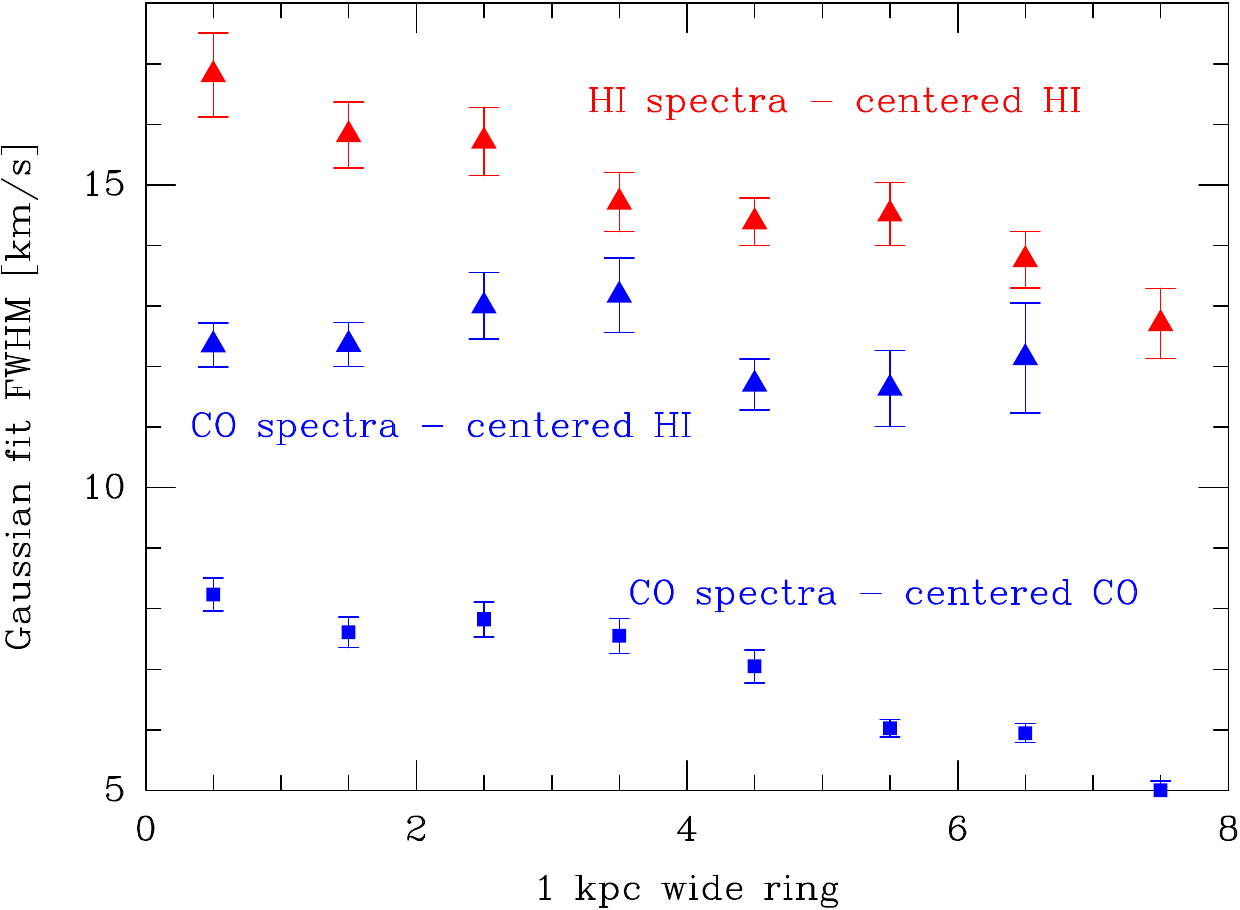}
    \caption{Evolution of the linewidth along the radius for the CO(2--1) and the \ion{H}{i} fits.}
    \label{fig:width}
  \end{figure}

  The middle panel of Fig.~\ref{fig:line-recenter} shows the CO spectra centered around the CO peak velocity detected above 4\,$\sigma$, which recovers less signal than the other CO centered spectra (left figure). The parameters of the corresponding fits show the intrinsic dispersion of the CO gas. However, since only a small fraction of M~33 has a CO brightness above 4\,$\sigma$, the "CO recentered CO" spectra cover relatively few lines of sight whereas recentering with the \ion{H}{I} velocity provides a virtually complete coverage and includes the regions with weak CO emission. The right panel shows the \ion{H}{I} recentered spectra, stacked in 1~kpc wide rings.  The half-power widths of all of these lines are plotted in Fig.~\ref{fig:width}.

  A clear decrease in the average CO and \ion{H}{i} linewidth with radius can be seen in Fig.~\ref{fig:width}. At all radii, the CO lines are narrower than the corresponding \ion{H}{i} lines despite both lines being recentered with respect to the \ion{H}{i}. This means that the disk becomes dynamically cooler with radius \citep{VanderKruit1982} and probably that the CO-\ion{H}{i} velocity dispersion decreases as well. While the former is expected, the latter is not necessarily because the outer disk is less efficient (because less gas-rich with longer rotation times) in circularizing velocities than the inner disk. The average lifetime of a GMC is only a small fraction of a rotation period.

  \section{Probability Distribution Functions} 

  Probability distribution functions (PDFs) of observables such as column density and temperature have been largely used for Galactic cloud studies \citep[e.g.][]{Lombardi2006,Kainulainen2009,Schneider2013}, and in numerical modeling \citep[e.g.][]{Federrath2008}. They have also been applied as an analytical tool for studying the intensity and temperature distribution from CO observations in galaxies \citep[M~51,][]{Hughes2013}.  We here produce a PDF of column density, derived from our CO observations of M~33. For that, we calculate the H$_2$ column density from the integrated CO intensity using the conversion factor 4~$\times$~10$^{20}$ $\Xunit$ \citep{Gratier2010a}. This is a straightforward approach to compare with Galactic observations \citep[e.g][]{Schneider2012, Russeil2013} and simulations \citep{Federrath2012} to interpret the physical origin of the features observed in a PDF.

\begin{figure}
    \centering
    \includegraphics[width=9cm]{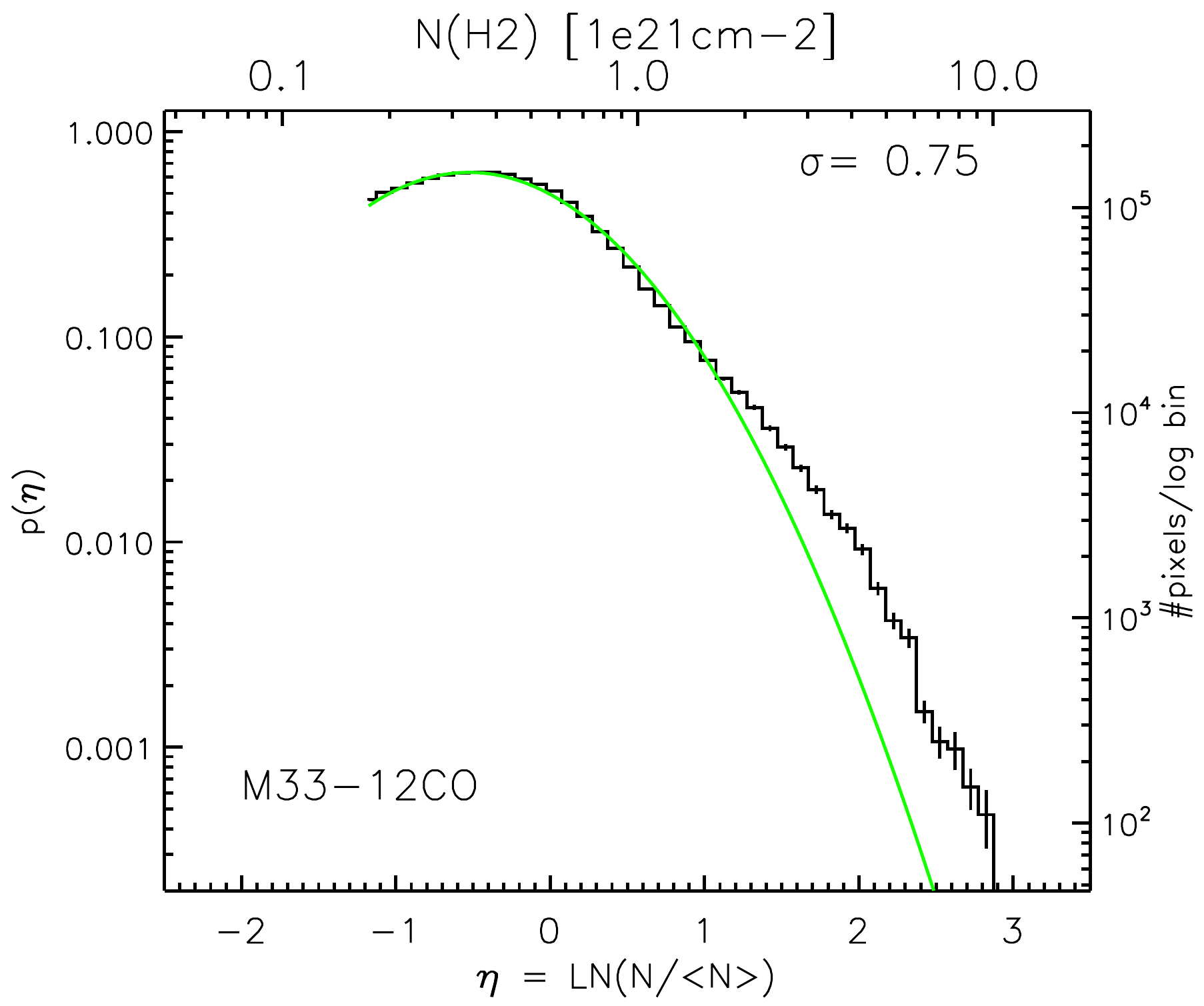}
    \caption{Probability distribution function of H$_2$ column density of M~33. The column density was derived from all pixels in the CO map from Fig.~\ref{fig:ico0K}, excluding only the noisy edges. The left y-axis gives the normalized probability $p(\eta)$, the right y-axis the number of pixels per log bin. The upper x-axis is in units of H$_2$ column density, and the lower x-axis is the logarithm of the normalized column density. The green curve indicates the fitted PDF. 
    The dispersion of the fitted PDF is indicated by $\sigma$.}
    \label{fig:PDF}
  \end{figure}

  All observed pixels from the map shown in Fig.~\ref{fig:ico0K} (only excluding the noisy edges seen in Fig.~\ref{fig:deltaico0K}) are considered and binned, and normalized to the average column density obtained from the same pixel statistic. The resulting PDF is shown in Fig.~\ref{fig:PDF}, expressed as a probability $p(\eta)$ (see also \citet{Federrath2008} for their definition of a 2D-PDF) with:
  \begin{equation}  
    \eta\equiv{\rm ln}\frac{N_\mathrm{H_2}}{\langle N_\mathrm{H_2} \rangle}  
  \end{equation}  
  In order to derive the characteristic properties of the PDF (width, peak, deviations(s) from the log-normal shape), we fit the log-normal function:
  \begin{equation}  
    p_\eta\,{\rm d}\eta=\frac{1}{\sqrt{2\pi\sigma^2}}{\rm exp}\Big[ -\frac{(\eta-\mu)^2}{2\sigma^2} \Big]{\rm d}\eta  
  \end{equation}  
  where $\sigma$ is the dispersion and $\mu$ is the mean logarithmic column density. We do this systematically by performing several fits on a grid of parameters for $\eta$ and $\mu$ and then calculate the positive and negative residuals. Because excess is expected to lie above the log-normal form, we select fits with the least negative residuals. We then determine the range of log-normality, when the difference between the model and $p_\eta$ is less than three times the statistical noise in $p_\eta$. 

  The PDF we obtain (Fig.~\ref{fig:PDF}) shows a clearly defined log-normal distribution for low column densities with a peak around 0.5~$\times$~10$^{21}$~cm$^{-2}$ and excess above $\sim$1.7~$\times$~10$^{21}$cm$^{-2}$. We emphasize that it is not possible to fit a much broader log-normal PDF in order to cover this higher column density range.  Assuming that the underlying property indeed has a log-normal distribution\footnote{If the density (or column density) is determined by a large number of independent random fluctuations, the quantity $\eta$=N/<N> is determined by their sum, and $p_\eta$ becomes a Gaussian distribution according to the central limit theorem \citep[see also][]{Vazquez1994,Federrath2010}}, it is important to fit a {\sl log-normal} PDF to the peak, the column densities left of the peak until the noise limit (indicated in Fig.~\ref{fig:PDF} as a dotted line), and the column densities right of the PDF peak. We then obtain a log-normal PDF with a width of 0.75 (in units of ln(N/$\langle$ N $\rangle$).

 The excess we observe is in the column density range  $\sim$1.7~$\times$~10$^{21}$~cm$^{-2}$ to $\sim$4~$\times$~10$^{21}$cm$^{-2}$. If a power-law is fit to the data in this column density range, then the slope is $s \approx - 2.4$, typical of the slopes of the power-law tails in Galactic molecular clouds \citep{Russeil2013,Schneider2013,Schneider2014} and attributed mainly to self-gravity. Recently, this interpretation was confirmed in analytic collapse models by \citet{Girichidis2014}. Other physical processes can affect the shape of the PDF. External compression such as expanding \ion{H}{ii}-regions leads to a broader PDF \citep{Schneider2013} and/or to a "double-peaked" PDF \citep{Schneider2012,Tremblin2014}. However, these detailed features in the PDF of Galactic clouds are diluted in the galaxy-wide PDF such as the one for M~33.

 The most straightforward explanation is that self-gravity is causing the excess at high column densities in M~33. In their study of M~51 at a comparable spatial resolution, \citet{Hughes2013} did not find such an excess in the CO intensities observed in M~51 (e.g. left panel of their Fig.~2).  However, we believe that this is due to their fitting procedure and that a break probably occurs close to log($I_\mathrm{CO}$) = 1.8. Numerical models \citep[e.g.][]{Girichidis2014} suggest that the power-law tail for molecular clouds is caused by self-gravity of the clumps and cores located inside the cloud, not necessarily by free-fall contraction of the whole cloud. Numerical hydrodynamical models of galactic disks come to different predictions for the PDF. \citet{Wada2007} find density PDFs (note that we observe the column density PDF) that resemble more log-normal shapes in their simulations including self-gravity and heating/cooling, while \citet{Dobbs2008} obtain a density PDF with a power-law tail in their SPH simulations including self-gravity and adiabatic two-phase gas. They conclude that both, agglomeration of small clouds and self-gravity produce GMCs in spiral galaxies.

\section{Conclusions}

  In this paper, the first complete map of the CO(2--1) emission in M~33 up to R$_{\mathrm{opt}}$ ($\sim$7~kpc) and the associated integrated intensity map are presented. The average noise level of 20.33~mK per 2.6$\kms$ channel and the angular resolution is 12$\arcsec$ or 49~pc at the assumed distance of M~33. 
  In addition to the CO(2--1) observations, CO(1--0) has been observed along a radial strip. Our main conclusions are the following:

  \begin{enumerate}

  \item  The total CO(2--1) luminosity is 2.8~$\times$~10$^7$ K$\kms$, corresponding to a molecular gas mass of 3.1~$\times$~10$^8$~M$_\sun$ assuming a conversion factor of $\ratio$ = 4~$\times$~10$^{20}\Xunit$, twice the classical Milky Way value. The uncertainty in the CO luminosity is dominated by calibration uncertainty of $\sim 15$\%. The surface density of molecular gas decreases exponentially with radius with a scale length of 2.1~kpc.

  \item  Down to a resolution of $\approx$~50~pc (GMC size scale), the velocity dispersion between atomic and molecular gas is very low. The CO(2--1) peak temperature follows the atomic gas peak brightness very closely, suggesting a tight connection between the atomic and molecular components..
  While shifting the CO(2--1) spectra by the velocity of the \ion{H}{i} peak (sect \ref{sec:recenter}) and stacking the spectra, the CO line is very narrow (12.4$\kms$) even summed over the whole disk.
  In addition, the linewidths of both components decrease with galactocentric distance, due to a lower molecular-atomic velocity dispersion and/or intrinsically narrower CO(2--1) lines.

  \item  The CO(2--1) emission observed towards \ion{H}{i}-poor regions of M~33 is at the level expected from the error beam pickup. Therefore, there is no evidence for molecular gas formation where the \ion{H}{i} peak temperature is below 10~K.  

  \item  The CO($\frac{2-1}{1-0}$) line ratio varies significantly over the disk but not in a regular fashion. The mean value is 0.8 which we apply to the ensemble of our data.
  
  \item  The Probability Density Function of the H$_2$ column density as traced by the CO emission exhibits a log-normal profile with a considerable excess in the high column density
regime, presumably due to the onset of gravitational contraction.

  \end{enumerate}

\begin{acknowledgements}
  We thank the IRAM staff for help provided during the observations and for data reduction.
  N.S. was supported by the ANR (\emph{Agence Nationale pour la Recherche}) project "STARFICH" number ANR-11-BS56-010.  
\end{acknowledgements}

\bibliographystyle{aa} 
\bibliography{papier}

\end{document}